\documentclass[%
 reprint,
 amsmath,amssymb,
 aps,
]{revtex4-2}
\usepackage{physics}
\usepackage{hyperref}
\usepackage{graphicx}% Include figure files
\usepackage{dcolumn}% Align table columns on decimal point
\usepackage{bm}
\usepackage{capt-of}

\newcommand{\neuo}{\widetilde{\chi}_{1}^{0}}
\newcommand{\neut}{\widetilde{\chi}_{2}^{0}}

\newcommand{\ch}{\widetilde{\chi}_{1}^{\pm}}
\DeclareMathOperator{\RKI}{\mathrm{RKI}}

\begin{document}

\preprint{APS/123-QED}

\title{Probing Compressed Mass Spectrum Supersymmetry at the LHC with the Vector Boson Fusion Topology}% Force line breaks with \\

\author{U. S. Qureshi}%
 \email{uqureshi@cern.ch}
\author{A. Gurrola}
 \email{alfredo.gurrola@vanderbilt.edu}
 
\affiliation{
Department of Physics and Astronomy,
Vanderbilt University, Nashville, TN, 37235, U.S.A.
}%

\author{A. Flórez}
  \email{ca.florez@uniandes.edu.co}
\affiliation{
Physics Department, Universidad de los Andes,
Bogotá, Colombia.
}%

\date{\today}% It is always \today, today,

\begin{abstract}
We present a phenomenology study probing pair production of supersymmetric charginos and neutralinos (``electroweakinos'') with the vector boson fusion (VBF) topology in proton-proton collisions at CERN's Large Hadron Collider (LHC). In particular, we examine the compressed-mass spectrum phase space that has been traditionally challenging due to experimental constraints. The final states considered have two jets, large missing transverse momentum, and one, two, or three light leptons. Different model scenarios are considered for the production and decays of the electroweakinos. A novel high-performance and interpretable sequential attention-based machine learning algorithm is employed for signal-background discrimination and is observed to significantly improve signal sensitivity over traditional methods. We report expected signal significances for integrated luminosities of $137$, $300$, and $3000$ $\textrm{fb}^{-1}$ corresponding to the current data acquired at the LHC, expectation for the end of Run 3, and the expectation for the high-luminosity LHC. Our methodology results in projected 95\% confidence level bounds that cover chargino masses up to 1.1 TeV in compressed-mass spectrum scenarios within the R-parity conserving minimal supersymmetric standard model. This parameter space, currently beyond the reach of ATLAS and CMS searches at the LHC, is traditionally challenging to explore due to significant Standard Model backgrounds and low signal cross-sections.
\end{abstract}

\maketitle

% \tableofcontents

% %Graphical abstract


% \begin{keyword}
% %% keywords here, in the form: keyword \sep keyword, up to a maximum of 6 keywords
% BSM physics \sep dark matter\sep muon $g-2$\sep $R_D$ and $R_{D^*}$ anomalies.

% %% PACS codes here, in the form: \PACS code \sep code

% %% MSC codes here, in the form: \MSC code \sep code
% %% or \MSC[2008] code \sep code (2000 is the default)

% \end{keyword}

%\tableofcontents



\section{Introduction}
\label{introduction}

The Standard Model (SM) of particle physics, has been a largely successful theory, accounting for numerous experimental findings involving strong, electromagnetic, and weak interactions over recent decades. The SM has been extensively tested at the CERN's Large Hadron Collider (LHC) as a theory describing elementary particles and their interactions below the \textrm{TeV} scale. However, as experiments probe higher and higher energies, observations that contradict the SM's predictions have begun to emerge. These observations include the origins of dark matter \cite{WMAP:2012nax, Planck:2018vyg, Bertone:2004pz}, electroweak symmetry breaking scales \cite{Branco:2011iw, Gori:2016zto, BhupalDev:2014bir, Liu:2023jbq}, baryon asymmetry \cite{Sakharov:1967dj, Dine:2003ax}, neutrino masses \cite{Kajita:2016cak}, the anomalous muon magnetic moment \cite{Muong-2:2023cdq, Muong-2:2024hpx}, and discrepancies in the $R_{(D)}$ and $R_{(D^{*})}$ ratios from $B$-meson decays \cite{BaBar:2013mob, Belle:2015qfa, Belle:2016ure, Belle:2016dyj}. Furthermore, there are theoretical problems with how and if gravity should be quantized, how gauge interactions can be unified, and how to explain divergences in the Higgs mass calculations \cite{Hosotani:2017krs, Arkani-Hamed:1998jmv}. Furthermore, the SM offers no explanation for fermion family replication \cite{Georgi:1974sy, Froggatt:1978nt, Babu:2009fd}, nor for the lack of CP violation in the strong sector \cite{tHooft:1976rip, Peccei:1977hh, Kim:1986ax}. As such, the SM can at best be considered a low-energy effective field theory approximating a more complete theory. This reinforces the expectation for beyond the SM (BSM) physics. 

% \iffalse
As a result, several theoretical models have been put forth to address the limitations of the SM over the past decade. Despite differing theoretical motivations and resulting implications, a common thread among these ideas is the introduction of new particles, that can be probed via proton-proton collisions at the  LHC. A myriad of ideas have been suggested to investigate BSM physics, driving a substantial amount of exploration at LHC. Said research has significantly limited the scope of theories and established exclusionary boundaries, extending to multi-\textrm{TeV} ranges for the masses of newly predicted particles within these theories. Possible reasons for the absence of evidence could be attributed to new particles masses being at the edge or too large to be produced at the LHC energies and likely with exceptionally low production rates. In the scenario where the masses of the new particles might be probed at the LHC but their production cross sections are small with respect to SM processes, a vast amount of data might be needed, together with advanced analysis techniques to enhance the probability of detection. Alternatively, it is conceivable that new physics diverges from the conventional assumptions made in many BSM theories and the associated explorations. As a result, these new physics phenomena could remain hidden in processes that have not yet been thoroughly examined.
% \fi

Supersymmetry (SUSY)~\cite{Ramond:1971gb, Ferrara:1974pu, Wess:1974tw, Chamseddine:1982jx, Barbieri:1982eh, Martin:1997ns} is a well-motivated theoretical framework that aims to address several unresolved questions in the SM, including the identification of the particle that constitutes dark matter (DM). The key premise of supersymmetry is to restore the symmetry between the fermionic matter and bosonic force carrier fields of the SM. SUSY models accomplish this by including superpartners of SM particles such that the spins of these superpartners differ by one-half unit with respect to corresponding SM partners. In R-parity conserving SUSY models, superpartners are pair-produced in high-energy proton-proton (pp) collisions at the LHC, and their decay chains end with a stable, electrically neutral supersymmetric particle, typically referred to in the literature as the lightest supersymmetric particle (LSP). In several SUSY models (including the one considered in this paper), the LSP and canonical DM candidate is the lightest neutralino $\neuo$, which is a superposition of the wino, bino, and higgsino fields that form the SUSY partners of the SM $W$, $\gamma/Z$, and $H$ fields respectively.

The ATLAS and CMS experiments at the LHC have an extensive program to search for SUSY; however, despite the compelling theoretical framework, no direct evidence for SUSY has yet been observed. The absence of any firm experimental evidence of SUSY signals at the LHC indicates two possible scenarios: either the SUSY particles are too heavy to be probed with the current LHC energies, or the SUSY particles are hidden in the phase space where sensitivity is limited due to experimental constraints. In this study, we probe the latter scenario in the so-called compressed-mass spectrum regime, where the mass differences $\Delta m$ between the LSP and other electroweak SUSY particles are small. Traditionally, in these so-called compressed spectrum scenarios, the momentum available to the co-produced SM particles, typically leptons, is small (``soft''). These very soft final states are challenging to reconstruct experimentally at the analysis level. Even in the case where these final states can be reconstructed, they are usually susceptible to large SM backgrounds, thus diminishing their discovery reach. 

In this paper we show that VBF tagging in combination with a novel machine learning approach to distinguish between signal and background can significantly alleviate these challenges. 
%It is worth pointing out that this is not merely an academic exercise. 
Compressed-mass spectra arise in several regions of the SUSY parameter space and lead to interesting physics scenarios. For example, in the stau-neutralino coannihilation region, the mass gap between the scalar superpartner of the $\tau$ lepton, the $\widetilde{\tau}$ and the $\neuo$ must be under 50 GeV to be consistent with the thermal dark matter relic density as observed by the Planck and WMAP Collaborations~\cite{Planck:2018vyg, WMAP:2012nax, Arnowitt:2008bz, Schwaller:2013baa}. Another example is the so-called natural SUSY models \cite{Feng:2013pwa} that solve the hierarchy problem with minimal fine-tuning. In these models, the bino, wino, and higgsino mass eigenstates are parameterized by the soft SUSY breaking terms $M_1$ and $M_2$, and the superpotential higgsino mass parameter $\mu$. 
%The origin of these terms is detailed in Section \ref{sec:model} but for now we note that such natural SUSY models impose constraints on higgsino masses and suggest that $\abs{\mu}$ be near the weak scale while $M_1$ or $M_2$ be larger \cite{Baer:2020sgm}. 
In such a scenario, the lightest SUSY particles are electroweakinos which are a triplet of higgsino-like states, such that the mass difference between them is small and determined by the difference between $M_1$ or $M_2$ and $\abs{\mu}$~\cite{Baer:2020sgm}.

\begin{figure}
    \centering
    \includegraphics[width=\linewidth]{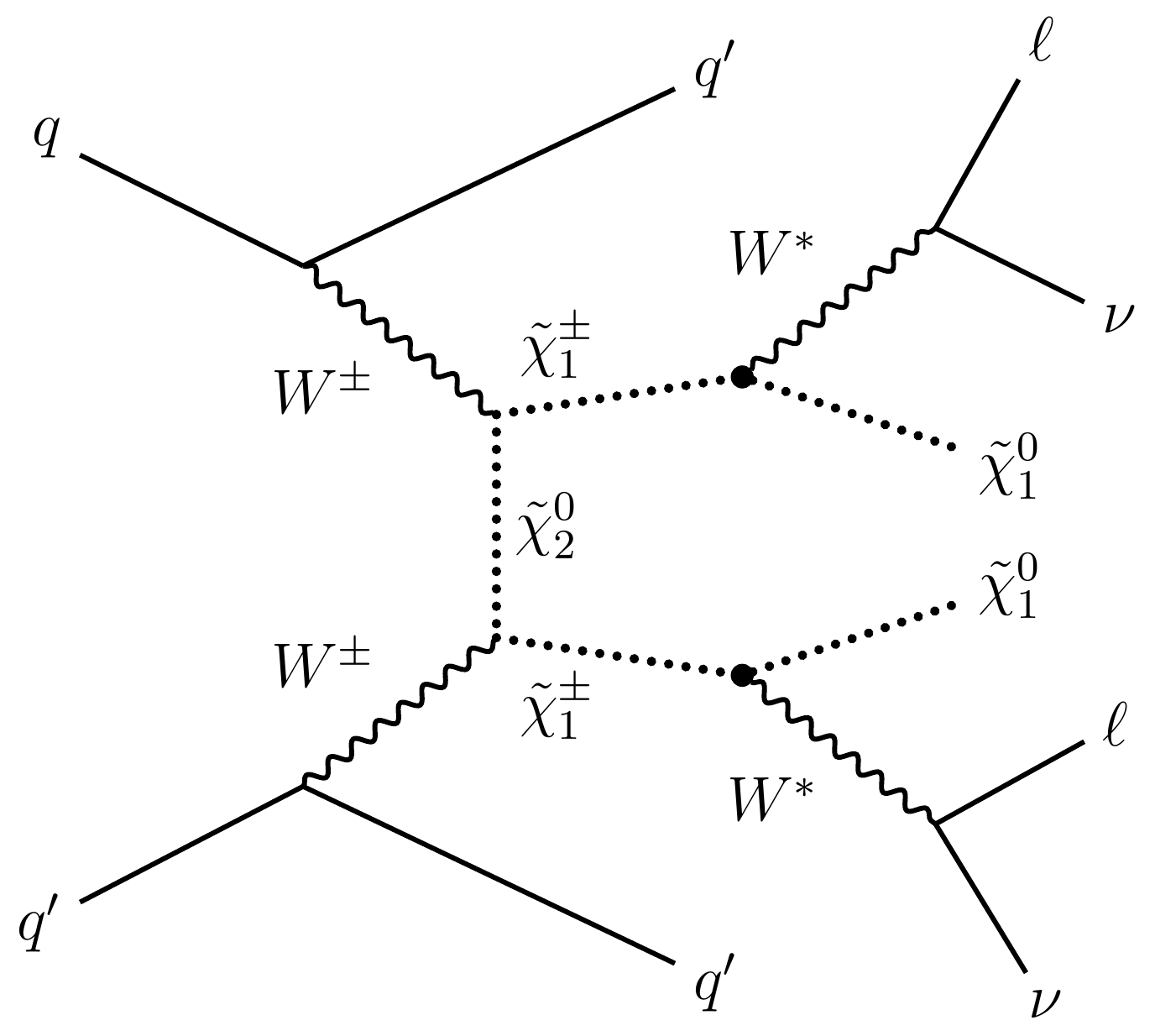}
    \caption{Representative Feynman diagram for VBF $t$-channel chargino-chargino pair production, and subsequent decay to a 2-lepton final state through a $W^*$ boson in the Wino-Bino $W^*/Z^*$ model scenario.}
    \label{fig:chchFeynmann}
\end{figure}

The search strategy employed in this paper utilizes the VBF dijet topology \cite{Dutta_2013, Delannoy:2013ata, Dutta:2014jda, Dutta:2013gga, Florez:2021zoo, Avila:2018sja, Arnowitt:2008bz} in which electroweak SUSY particles are pair-produced in association with two distinctive, energetic jets. Subsequent decays are then considered using different model scenarios detailed in Section \ref{sec:sims}, but all end in final states with two VBF-tagged jets, large missing transverse momentum, and either zero, one, two, three, or four leptons. Figure \ref{fig:chchFeynmann} shows representative Feynman diagrams for our signal processes considering the so-called ``Wino-Bino $W^*/Z^*$ model'' scenario, where the $\widetilde{\chi}_{1}^{0}$ is Bino, the $\widetilde{\chi}_{1}^{\pm}$ and $\widetilde{\chi}_{2}^{0}$ are Wino, the $\widetilde{\chi}_{1}^{\pm}$ mass $m(\widetilde{\chi}_{1}^{\pm})$ is equal to $m(\widetilde{\chi}_{2}^{0})$ since the $\widetilde{\chi}_{1}^{\pm}$ and $\widetilde{\chi}_{2}^{0}$ belong to the same group multiplet, and the chargino/neutralino decays proceed via off-mass shell W$^{*}$/Z$^{*}$ bosons. As  shown in Figure~\ref{fig:chchFeynmann}, different combinations of electroweakino pair production lead to final states of varying lepton multiplicity. For example, pair production of the lightest chargino ($\widetilde{\chi}_{1}^{\pm}$) can lead to two leptons in the final state, while production of $\widetilde{\chi}_{1}^{\pm}$ and the second lightest neutralino ($\widetilde{\chi}_{2}^{0}$) can lead to three leptons in the final state. 

The VBF topology has proved to be a powerful experimental tool for searches at the LHC due to the remarkable control over SM backgrounds, while also creating a kinematically boosted topology \cite{CMS:2016ucr, Florez:2016uob, CMS:2015jsu}. This makes it possible to reconstruct and identify the aforementioned soft decay products typical of compressed mass spectra. Although the topologies considered in this analysis result in final states with up to four leptons, the decay products have a peak transverse momentum $\expval{p_T} \sim \mathcal{O}(\Delta m) $, which makes it difficult to experimentally reconstruct and identify all of them, especially when $\Delta m  < 5$ GeV and thus the lepton $p_{\textrm{T}}$ are too soft to effectively reconstruct those particles at the LHC. As will be described later in Section \ref{sec:ML}, the kinematics of the said soft lepton plays a non-trivial role in the signal-background discrimination process, and thus also the projected high-luminosity LHC (HL-LHC) discovery reach. This is particularly important in the case of the HL-LHC scenario where the usage of such soft lepton kinematics is likely challenging due to large number of secondary pp interactions (pileup).  We note that pileup mitigation has been well-studied at ATLAS and CMS \cite{CMS-PAS-FTR-13-014}, and therefore in this paper we attempt to provide reasonable expectations for the performance of particle identification and reconstruction. We conservatively take into account degradation in particle identification efficiencies by considering the case of 140 average pileup interactions, discussed further in Section \ref{sec:results}. 

A key component of this study is also the development of an analysis strategy utilizing a novel deep-learning architecture. This algorithm uses a sequential attention mechanism to choose a subset of semantically meaningful features to process at each decision step. Instance-wise feature selection enables efficient learning as the model capacity is fully used for the most salient features. This results in better performance and significantly enhances explainability and interpretability. This method has been shown to outperform traditional machine learning-based approaches for event classification like boosted decision trees and feed-forward multi-layer perceptions that found use in high-energy physics. The event classifier's output is employed to conduct a profile-binned likelihood fit, which is used to determine the overall signal significance for each model examined in the analysis. Our findings indicate that this algorithm significantly enhances signal significance.

The rest of this paper is structured as follows.  Section~\ref{sec:exp} provides an overview of current results from searches for Supersymmetry at the LHC. Section~\ref{sec:sims} explains how the Monte
Carlo simulation samples are produced for this study. In Section~\ref{sec:ML} we discuss the motivation and details of our machine learning workflow, and in Section~\ref{sec:results}, the main results are presented. We conclude with a short discussion in Section~\ref{sec:discussion}. 
%Appendix \ref{sec:model} provides salient details on the minimal supersymmetric standard model.

\section{Experimental Considerations}\label{sec:exp}
\begin{figure}
    \centering
    \includegraphics[width=\linewidth]{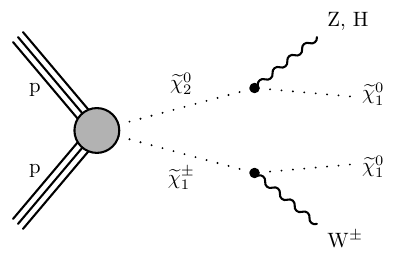}
    \caption{Representative Feynmann diagram from \cite{CMS:2024gyw} for the production of $\neut$ and $\ch$, with the $\neut$ decaying to either a $Z$ or $H$ boson and a $\neuo$, and the $\ch$ decaying to a $W^\pm$ boson and a $\neuo$ in the Wino-Bino model scenario.}
    \label{fig:CMS-SUS-21-008_Figure_001}
\end{figure}

The ATLAS and CMS collaborations at CERN have conducted various searches for SUSY particles over the past fifteen years. These searches utilized pp collisions at center-of-mass energies of $\sqrt{s} = 7$, $8$ and $13$ \textrm{TeV}. As a result, we have stringent bounds on the masses of the colored SUSY sector. They exclude, at the 95\% confidence level, gluino ($\widetilde{g}$) masses up to 2.2 TeV and squark ($\widetilde{q}$) masses up to 1.8 GeV, depending on the model \cite{CMS:2023zuu, CMS:2023mny, CMS:2023xlp}. 
However, bounds on the electroweak sector (neutralinos and charginos) are less constrained, since, expectedly, these particles suffer from smaller electroweak production rates. Constraints on compressed $\Delta m$ SUSY scenarios were first set by the LEP experiments~\cite{Heister:2001nk,Abdallah:2003xe,Achard:2003ge,Abbiendi:2003ji}, where the lower bound on direct $\widetilde{\chi}_{1}^{\pm}$ or $\widetilde{\chi}_{2}^{0}$ production was found to be $m(\widetilde{\chi}_{1}^{\pm}) \approx 75$-$92.4$ GeV (103.5 GeV), depending on the mass difference $\Delta m(\widetilde{\chi}_{1}^{\pm}, \widetilde{\chi}_{1}^{0})$. Specifically, for $\Delta m(\widetilde{\chi}_{1}^{\pm}, \widetilde{\chi}_{1}^{0}) < 3$ GeV, the bound is between 75 and 92.4 GeV, while for $\Delta m > 3$ GeV, it is approximately 103.5 GeV. At the LHC, CMS/ATLAS searches for electroweakinos have mostly focused on pair production from Drell-Yan-like $qq^{\prime}$ annihilation processes such as the one depicted in Figure \ref{fig:CMS-SUS-21-008_Figure_001}. Subsequent decays and final states depend on the particular search strategy and model, which we summarize in the remainder of this section.

The CMS search in Ref.~\cite{CMS:2020bfa} considers final states with an opposite-sign same-flavor lepton pair and $p_T^{\mathrm{miss}}$. The analysis targets the 
decay of $\widetilde{\chi}_{1}^{\pm}$ to a W boson, and the decay of $\widetilde{\chi}_{2}^{0}$ to either a Z or Higgs boson. Therefore, the analysis strategy involves either the reconstruction of a lepton pair whose invariant mass is consistent with the Z boson mass, a kinematic edge in the invariant mass distribution of the lepton pair, or the nonresonant production of two leptons. In that analysis, charginos and neutralinos with masses up to about 750 GeV are excluded at the 95\% confidence level, assuming a massless LSP. On the other hand, the bounds on the $\widetilde{\chi}_{1}^{\pm}$ and $\widetilde{\chi}_{2}^{0}$ mases do not exceed the LEP bounds of about 100 GeV for mass gaps below 50 GeV. Meanwhile, the CMS search in Ref.~\cite{CMS:2021few} considers final states with one lepton, two $b$ jets, and missing transverse momentum ($p_{\textrm{T}}^{\textrm{miss}}$), assuming the $\widetilde{\chi}_{1}^{\pm}$ decays to a W boson and the LSP and the $\widetilde{\chi}_{2}^{0}$ decays to a Higgs boson and the LSP. In this analysis, charginos and neutralinos with masses up to 820 GeV are excluded at 95\% confidence level when the LSP mass is small. However, similar to Ref.~\cite{CMS:2020bfa}, the bounds do not exceed the LEP bounds for mass gaps below 50 GeV. The CMS search in Ref.~\cite{CMS:2022vpy} considers a topology with four $b$ jets and $p_{\textrm{T}}^{\textrm{miss}}$, targeting $\widetilde{\chi}_{2}^{0}\widetilde{\chi}_{2}^{0}$ production and subsequent decays to a pair of H bosons that each subsequently decay through H$\to b\bar{b}$. Furthermore, the CMS search in Ref.~\cite{CMS:2022sfi} considers final states with multiple jets and $p_T^{\mathrm{miss}}$ targeting the hadronically decaying $WW$, $WZ$, and $WH$ topologies. In these analyses, Wino-like (Higgsino-like) particles with masses up to 960 (650) GeV are excluded, assuming a massless LSP. Similar to the Refs.~\cite{CMS:2020bfa,CMS:2021few,CMS:2020bfa}, the bounds do not exceed the LEP bounds for mass gaps below 50 GeV.

In general, as noted above, the bounds in the compressed-spectrum regime are much lower, and thus require targeted and innovative search methods. For example, the searches in Refs.~\cite{CMS:2021edw,CMS:2021cox} consider chargino-neutralino production in a compressed mass region where the mass gaps are less than 50 GeV, and thus the decays proceed through W and Z bosons that are off-mass shell. In this scenario, the decay products have low momentum, and thus the analysis strategy utilizes a large missing momentum requirement (e.g., $p_{\textrm{T}}^{\textrm{miss}} > 125$ GeV in Ref.~\cite{CMS:2021edw}) to boost the system and promote larger signal acceptance from the lepton kinematic requirements, combined with innovative soft-lepton reconstruction and identification algorithms using machine learning, trained for $p_{\textrm{T}}(\ell) < 10$ GeV, in order select two, three, or four soft leptons to further mitigate SM backgrounds. Reference~\cite{CMS:2021cox} additionally makes use of a parameteric neural network, using the reconstructed masses of various lepton plus $p_{\textrm{T}}^{\textrm{miss}}$ combinations, for signal-background discrimination in models with large backgrounds. These innovative methods pay huge dividends at low $\Delta m$. Exclusion limits at 95\% confidence level are set on chargino and neutralino masses up to 275 GeV for a mass difference $\Delta m$ of 10 GeV in the Wino-Bino model, and up to 205 (150) GeV for a mass difference of 7.5 (3) GeV in the Higgsino model. %The two same-sign leptons, or three or more leptons, and $p_T^{\mathrm{miss}}$ CMS search \cite{CMS:2021cox} focuses on final states with either two light same-sign leptons or three or more leptons including $\tau_h$, and $p_T^{\mathrm{miss}}$. The analysis targets the $WZ$, $WH$, $ZZ$, $HZ$, and $HH$ topologies. A parametric neural network is used for signal-background discrimination in models with large backgrounds. Depending on the model hypotheses, charginos and neutralinos with masses up to values between 300 and 1450 GeV are excluded at the 95\% confidence level.

\begin{figure}
    \centering
    \includegraphics[width=\linewidth]{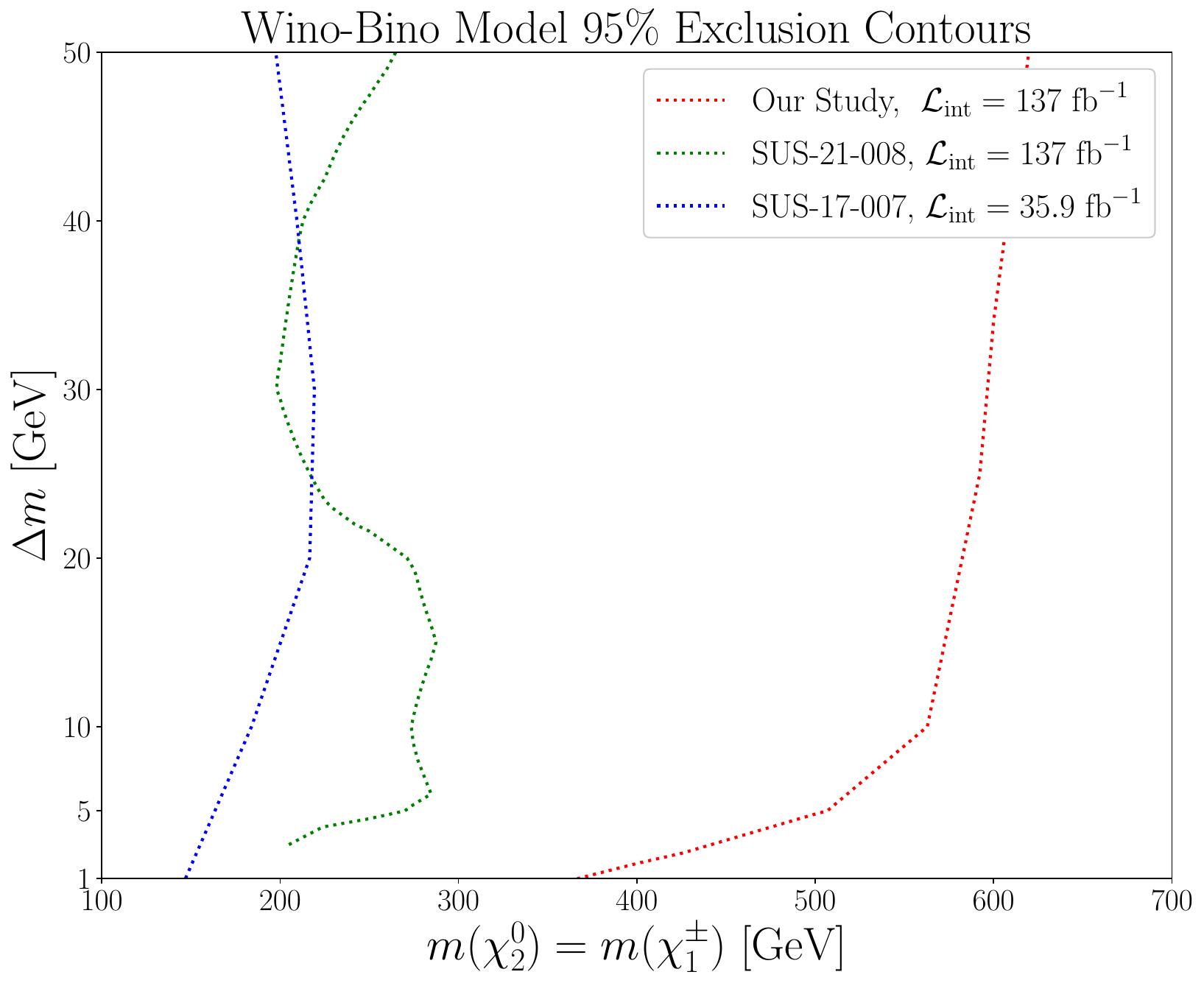}
    \caption{Wino-bino model exclusion contours targeting the $WZ$ topology for the compressed-mass region studied in this paper (red-dashed), the CMS combined search SUS-21-008 \cite{CMS:2024gyw}, and the CMS vector boson fusion search SUS-17-007 \cite{CMS:2019san}.}
    \label{fig:CMS-SUS-21-008_Figure_012-b}
\end{figure}

Combined results of several of the aforementioned searches for Drell-Yan production of electroweak SUSY particles are presented in \cite{CMS:2024gyw}. Those results utilize pp collision data collected during 2016-2018, which amounts to an integrated luminosity of about $137$ fb$^{-1}$. Figure \ref{fig:CMS-SUS-21-008_Figure_012-b} shows the exclusion contours from this analysis of combined results (represented as SUS-21-008 in the legend) for the Wino-bino model where the $\widetilde{\chi}_{1}^{\pm}$ ($\widetilde{\chi}_{2}^{0}$) decays via off-mass shell W (Z) bosons, for the compressed-mass spectrum region targeted in this study where $\Delta m = m(\widetilde{\chi}_{1}^{\pm}) - m(\widetilde{\chi}_{2}^{0}) \in \{ 1, 50\}$ GeV. For a compressed mass spectrum scenario in which $\Delta m = 10$ $(30)$ GeV and in which $\widetilde{\chi}_{1}^{\pm}$ and $\widetilde{\chi}_{2}^{0}$ branching fractions to
W and Z bosons are 100\%, $\widetilde{\chi}_{1}^{\pm}$ and $\widetilde{\chi}_{2}^{0}$ masses up to 260 (200) GeV are excluded at 95\% CL. On the other hand, no bounds from this study exist for $\Delta m < 3$ GeV. To facilitate a comprehensive analysis, benchmark masses in our study are considered down to 200 GeV for the neutralinos and charginos.

The use of the VBF signature as a way to probe SUSY models has been proposed several times, and this approach was explored experimentally by the CMS experiment in Ref~\cite{CMS:2019san} using the one- and zero-lepton topologies. The results of that analysis are shown in Figure~\ref{fig:CMS-SUS-21-008_Figure_012-b} and is represented as SUS-17-007 in the legend. For a compressed mass spectrum scenario in which the $\ch$ and $\tilde{\chi}_{2}^{0}$ decays proceed via a light slepton and the mass difference between the lightest neutralino and the mass-degenerate particles $\ch$ and $\tilde{\chi}_{2}^{0}$ is 1 (30) GeV, that analysis provides the most stringent lower limits of 112 (215) GeV on the $\ch$ and $\tilde{\chi}_{2}^{0}$ masses. However, the analysis in Ref~\cite{CMS:2019san} only utilizes about 36 fb$^{-1}$ of data, and a similar analysis with the 137 fb$^{-1}$ used by SUS-21-008 does exist. However, even with only 26\% of the integrated luminosity, the VBF analysis provides competitive sensitivity and motivates us to study the projected reach of such an analysis, using an improved analysis strategy detailed in the remainder of this paper.

% \begin{figure}
%     \centering
%     \includegraphics[width=\linewidth]{figures/CMS-SUS-21-008_Figure_012-a.pdf}
%     \caption{Wino-bino model exclusion contours from \cite{CMS:2024gyw} targeting the $WZ$ topology for the entire parameter space.}
%     \label{fig:CMS-SUS-21-008_Figure_012-a}
% \end{figure}

\section{Samples and Simulation}\label{sec:sims}

\begin{figure}
    \centering
    % \begin{minipage}{0.475\textwidth}
    \includegraphics[width=\linewidth]{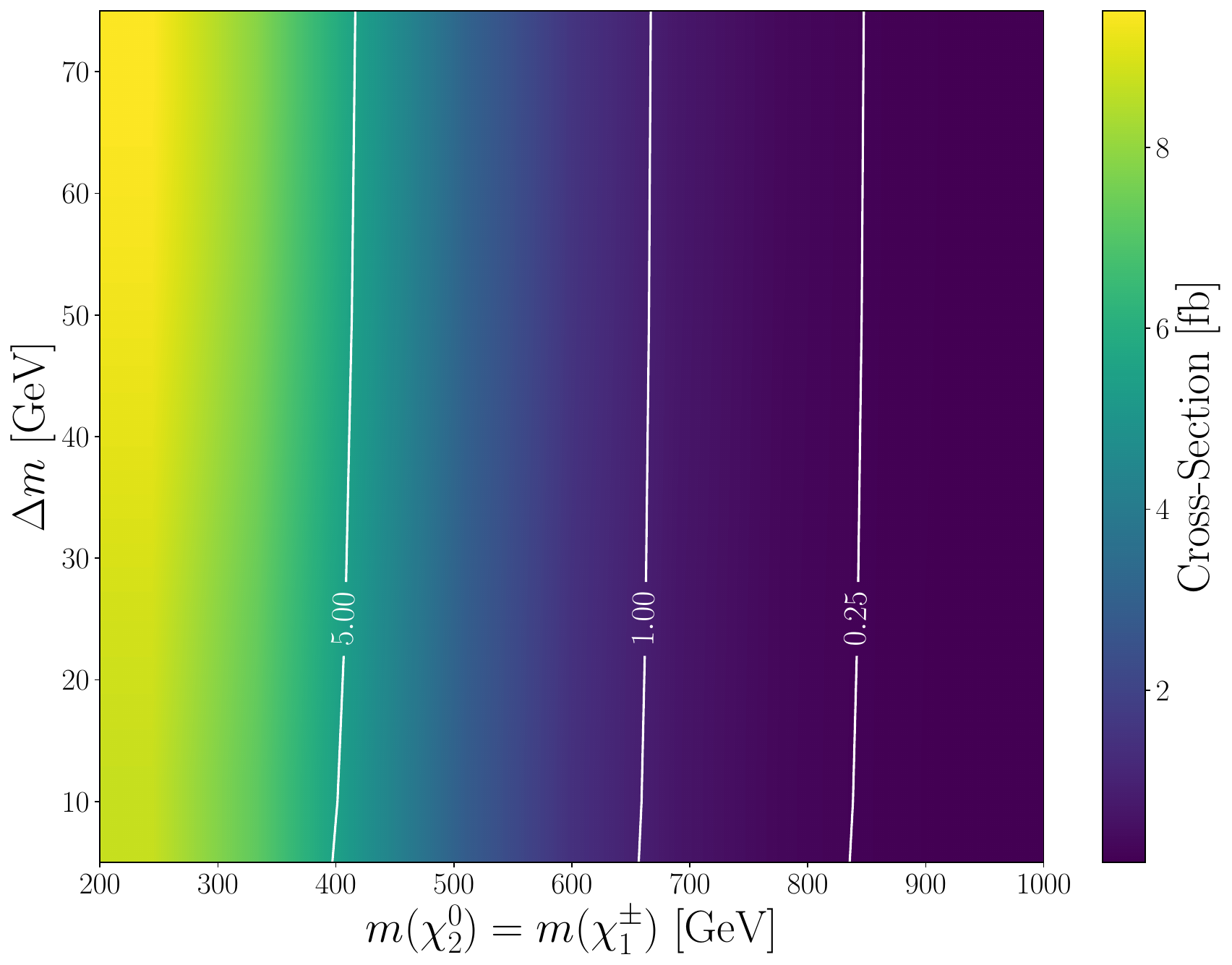}
        \captionof{figure}{Projected cross-section in femtobarns as a function of the $\ch$ and $\neut$ masses and the mass difference $\Delta m$ for the Wino-Bino $W^*/Z^*$ scenario.}
        \label{fig:CSBW}
    % \end{minipage}
    % \hspace{0.025\textwidth}
    % \begin{minipage}{0.475\textwidth}
    % \includegraphics[width=\linewidth]{figures/CS_LS.pdf}
    %         \captionof{figure}{Projected cross-section in femtobarns as a function of the $\ch$ and $\neut$ masses and the mass difference $\Delta m$ for the Wino-Bino light-slepton scenario.}
    %         \label{fig:CSLS}
    % \end{minipage}
    % \label{fig:enter-label}
\end{figure}

Signal and background samples are generated with \texttt{MadGraph5\_aMC 3.2.0} \cite{Alwall:2014hca, Frederix:2018nkq} considering $pp$ beams colliding with a center-of-mass energy of $\sqrt{s} = 13.6$ {TeV}, using the $\mathrm{R}$-parity conserving minimal supersymmetric standard model. The \texttt{NNPDF3.0\_NLO} \cite{NNPDF:2014otw} set of parton distribution functions
(PDFs) is used for all event generation.  Parton-level events are then interfaced with \texttt{Pythia 8.2.30} \cite{Sjostrand:2014zea, Sjostrand:2007gs} to account for parton showering and hadronization. Finally, we use \texttt{Delphes
3.4.1} \cite{deFavereau:2013fsa} to simulate detector effects using the CMS detector parameters for particle identification and reconstruction. The MLM algorithm is used for jet matching and jet merging. The parameters \texttt{xqcut} and \texttt{qcut} of the MLM algorithm are set to 30 and 45 respectively to ensure continuity of the differential jet rate as a function of jet multiplicity.

% \begin{figure}
%     \centering
%         \includegraphics[width=\linewidth]{figures/CS_BW.pdf}
%     \caption{Projected cross-section in femtobarns as a function of the $\ch$ and $\neut$ masses and the mass difference $\Delta m$ for the Wino-Bino $W^*/Z^*$ scenario.}
%     \label{fig:CSBW}
% \end{figure}

% \begin{figure}
%     \centering
%         \includegraphics[width=\linewidth]{figures/CS_LS.pdf}
%     \caption{Projected cross-section in femtobarns as a function of the $\ch$ and $\neut$ masses and the mass difference $\Delta m$ for the Wino-Bino light-slepton scenario.}
%     \label{fig:CSLS}
% \end{figure}

Signal samples are simulated considering chargino-neutralino pair-production with two associated jets: $pp \to \ch \ch j j$, $pp \to \widetilde{\chi}_{1}^{\pm}\widetilde{\chi}_{1}^{\mp} j j$, $pp \to \neuo \ch j j$, $pp \to \neut \ch j j$,  $pp \to \neut \neut j j$, $pp \to \neuo \neut j j$, and $pp \to \neuo \neuo j j$. We will refer to this process as $\widetilde{\chi}_{m}\widetilde{\chi}_{n}jj$. The chargino-neutralino
production is followed by $\neut \to \neuo \ell^\pm \ell^\mp $ and $\ch \to \neuo \ell^\pm \nu_\ell$ decays where $\ell =\{e,\mu,\tau\}$. Final states with a hadronically decaying $\tau$ lepton $\tau_h$ are not considered due to known experimental difficulties reconstructing soft genuine $\tau_h$ candidates with $p_{T} < 20$ GeV. In particular, they do not produce a narrow energy flow in the detector, making them difficult to distinguish from light-quark or gluon jets. 

Figures \ref{fig:CSBW} shows the signal production cross-sections as a function of $m(\widetilde{\chi}_{1}^{\pm})$ and the mass difference $\Delta m = m(\widetilde{\chi}_{1}^{\pm}) - m(\widetilde{\chi}_{1}^{0})$ for the Wino-Bino $W^*/Z^*$ model. All cross sections in this analysis are obtained requiring the following kinematic criteria on leptons $\ell$ and jets $j$ at the generator level: $|\eta(\ell)| < 2.5$, $p_{T}(j) > 20$ GeV, and $\abs{\eta(j)} < 5$. To suppress non-VBF contributions, additional generator-level cuts of $\abs{\Delta \eta(jj)} > 2.5$ and $m(jj)>200$ GeV are imposed. The cross section is about 9 fb at $m(\widetilde{\chi}_{1}^{\pm}) = 200$ GeV and 1 fb at $m(\widetilde{\chi}_{1}^{\pm}) = 650$ GeV. 

\begin{figure}
    \centering
    \includegraphics[width=\linewidth]{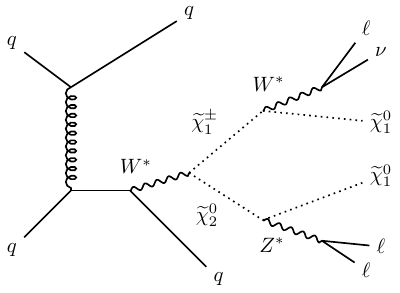}
    \caption{Representative Feynman diagram for mixed electroweak-QCD chargino-neutralino pair production with two associated jets, and subsequent decay to a 3-lepton final state through $Z^*/W^*$ bosons in the Wino-Bino $W^*/Z^*$ model scenario.}
    \label{fig:QCDfymn}
\end{figure}

We note that the two jets in $\widetilde{\chi}_{m}\widetilde{\chi}_{n}jj$ production can also arise from QCD couplings in diagrams such as in Figure \ref{fig:QCDfymn}, and interfere with the pure electroweak VBF processes shown in Figure 1. The interference between these diagrams can affect the total cross-section and the kinematics of interest for these studies, in particular the dijet invariant mass $m(jj)$.  We study these potential mixing effects by utilizing three sets of signal samples per $\{ m(\ch), \Delta m \}$ hypothesis:  (i) pure electroweak production of $\widetilde{\chi}_{m}\widetilde{\chi}_{n}jj$, whose cross section we denote as $\sigma_{\textrm{QCD}=0}$; (ii) $\widetilde{\chi}_{m}\widetilde{\chi}_{n}jj$ production of order $\alpha_{s}^{2}\alpha_{EW}^{2}$, whose cross section we denote as $\sigma_{\textrm{QCD}=2,\textrm{QED}=2}$; (iii) production of $\widetilde{\chi}_{m}\widetilde{\chi}_{n}jj$ which is inclusive in $\alpha_{s}$ and $\alpha_{EW}$, whose cross section we denote as $\sigma_{\textrm{incl}}$. We observe that $\sigma_{\textrm{incl}}$ is not equal to the sum of $\sigma_{\textrm{QCD}=0}$ and $\sigma_{\textrm{QCD}=2,\textrm{QED}=2}$. This behavior is suggestive of important interference effects between the two processes. In order to further illustrate this mixing effect, we define the relative kinematic interference (RKI) functional in Equation \ref{eq:RKI}. 

% \begin{widetext}
\begin{equation}
    \begin{aligned}
        \RKI\left[m(jj)\right] =&\frac{1}{\sigma_{\text{incl}}}\bigg[ \dv{\sigma_{\text{incl}}}{m(jj)}-\\& \left(\dv{\sigma_{\text{QCD=0}}}{m(jj)} + \dv{\sigma_{\text{QCD=2, QED=2}}}{m(jj)}\right)\bigg]
    \end{aligned}
    \label{eq:RKI}
\end{equation}
% \end{widetext}

\begin{figure}
    \centering
    \includegraphics[width=\linewidth]{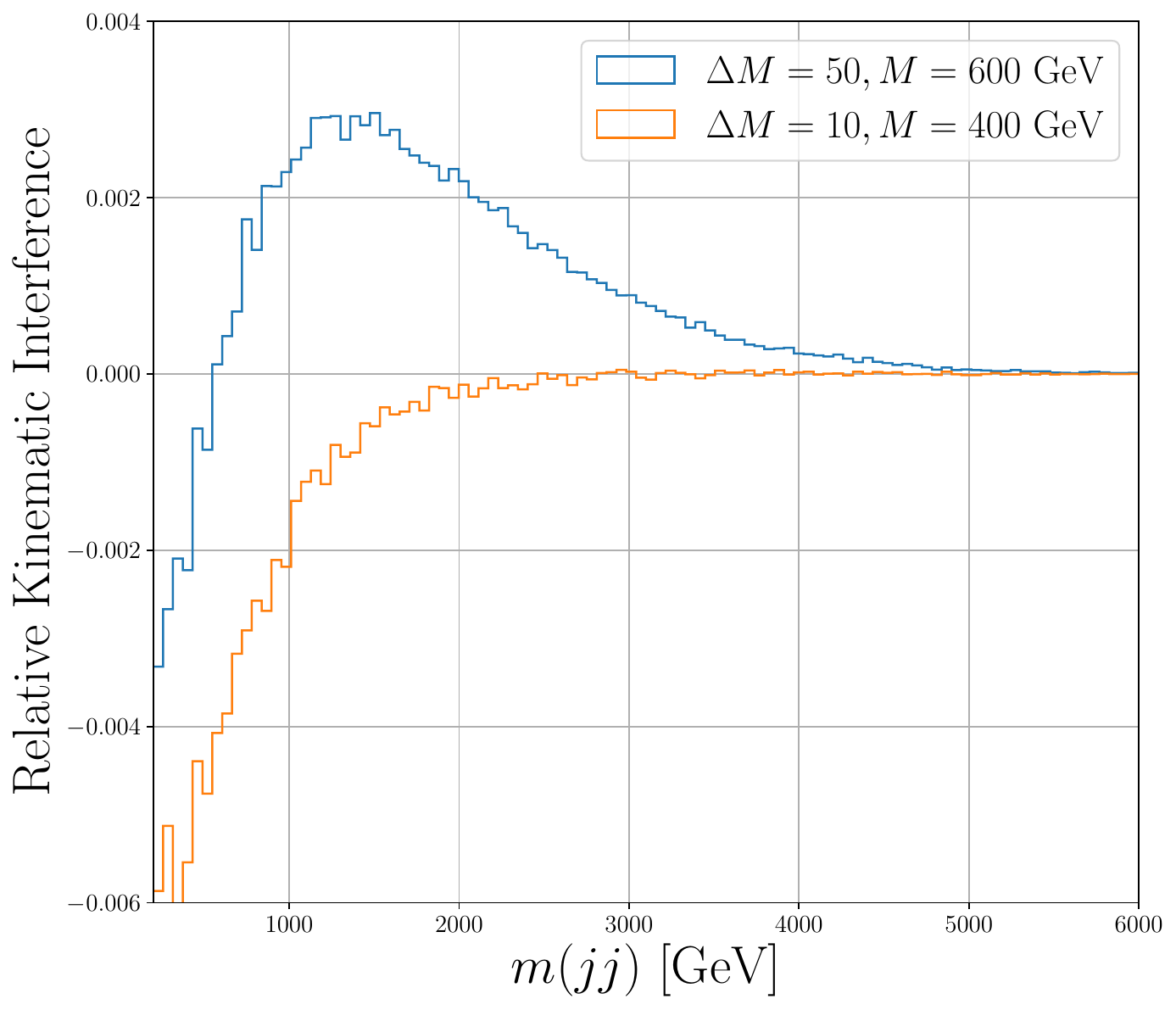}
    \caption{The relative kinematic interference (RKI), as a function of the reconstructed dijet invariant mass for a benchmark $m(\ch) = 600$ GeV, $\Delta m = 50$ Wino-Bino $W^*/Z^*$ scenario.}
    \label{fig:RKI}
\end{figure}

Figure \ref{fig:RKI} shows the RKI as a function of the reconstructed dijet invariant mass for two benchmark signal scenarios with $\{ m(\ch),\Delta m \} = \{ 400,10\}$ and $\{ 600,50\}$. The interference effect depends on both $m(\ch)$ and $\Delta m$, but is not negligible and can be up to about 20\% for the $\{ m(\ch),\Delta m \}$ values and ``ino’’ composition considered in this study. We note that this effect is smaller for larger values of $\{ m(\ch),\Delta m \}$, since $\sigma_{\textrm{QCD}=0} \gg \sigma_{\textrm{QCD}=2,\textrm{QED}=2}$. In the studies presented in the rest of this paper, we consider only the simulated signal samples that are produced inclusive in $\alpha_{s}$ and $\alpha_{EW}$ (i.e., the ones with $\sigma_{\textrm{incl}}$), since this ensures gauge invariant predictions.

For signal production, we consider two distinct SUSY models, the ``Wino-Bino $W^*/Z^*$'' and the ``Wino-Bino Light Slepton'' scenarios, which are detailed in the remainder of this section. 

%\subsection{Wino-Bino $\mathbf{W^*/Z^*}$ Scenario}

In the Wino-Bino $W^*/Z^*$ scenario, the $\neuo$ is purely bino and the LSP. The $\ch$ and $\neut$ are purely wino, mass degenerate (i.e., $m(\ch) = m(\neut)$), and the NLSPs. The three sleptons $(\widetilde{e}, \widetilde{\mu}, \widetilde{\tau})$ are mass degenerate, completely left-handed, and much heavier than the NLSPs. Therefore, the only allowed decay modes for the $\ch$ and $\neut$ are via virtual $W$ and $Z$ bosons. Hence, the following branching fraction ratios are unity: $\mathcal{B}(\ch \to \neuo W^{\pm,*} ) = 1$, $\mathcal{B}(\neut \to \neuo Z^{*} ) = 1$. 
%\begin{align}
%    \mathcal{B}(\ch \to \neuo W^{\pm,*} ) &= 1\\
%    \mathcal{B}(\neut \to \neuo Z^{*} ) &= 1
%\end{align}
The $W^{*}$ and $Z^{*}$ bosons subsequently decay to leptons. Figure \ref{fig:chchFeynmann} shows a representative Feynmann diagram for the production and decay of two charginos in this scenario. 

\begin{figure}
    \centering
    \includegraphics[width=\linewidth]{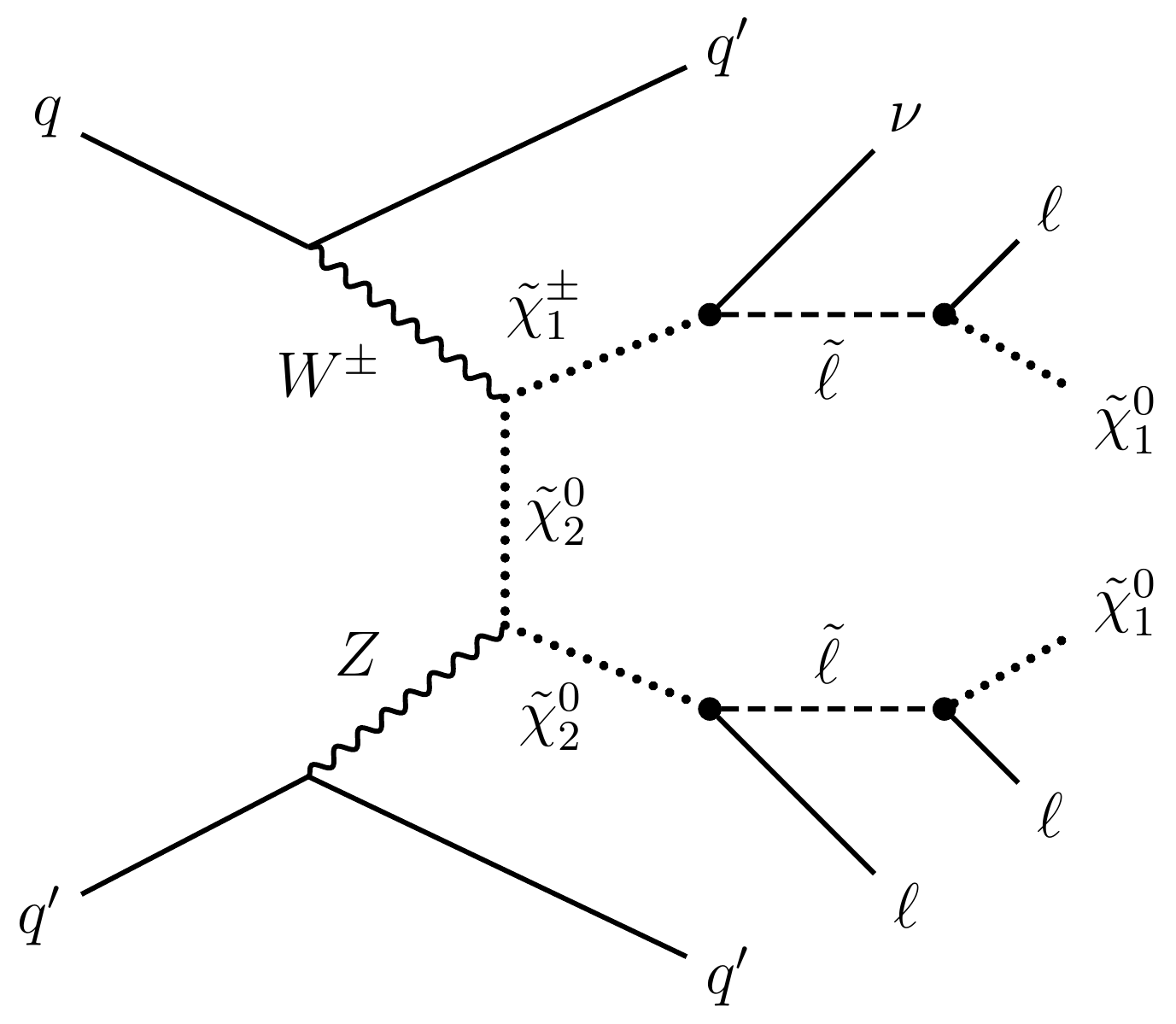}
    \caption{Representative Feynman diagram for VBF $t$-channel chargino-neutralino pair production, and subsequent decay to a 3-lepton final state through sleptons in the Wino-Bino light slepton model scenario.}
    \label{fig:slepton}
\end{figure}

%\subsection{Wino-Bino Light Slepton Scenario}
The Wino-Bino Light Slepton scenario is identical to the previous one except it differs in the slepton masses, thereby enabling the $\ch \to \widetilde{\ell} \nu_\ell $ and $\neut \to  \widetilde{\ell}^\pm \ell^\mp $ decay modes. In particular, the slepton masses are calculated as the average value of the $\neut$ and $\neuo$ masses: $m(\widetilde{\ell}) = \frac{m(\neut)+m(\neuo)}{2}$. 
%\begin{align}
%   m(\widetilde{\ell}) &= \frac{\neut+\neuo}{2}
%\end{align}
The branching fraction ratios for decays in this scenario are $\mathcal{B}(\ch \to \widetilde{\ell} \nu_\ell ) = \frac{1}{3}$, $\mathcal{B}(\neut \to  \widetilde{\ell}^\pm \ell^\mp ) = \frac{1}{3}$, and $\mathcal{B}(\widetilde{\ell} \to \neuo \ell ) = 1$, 
%\begin{align}
%    \mathcal{B}(\ch \to \widetilde{\ell} \nu_\ell ) &= \frac{1}{3}\\
%    \mathcal{B}(\neut \to  \widetilde{\ell}^\pm \ell^\mp ) &= \frac{1}{3}\\
%    \mathcal{B}(\widetilde{\ell} \to \neuo \ell ) &= 1
%\end{align}
where $\ell$ is one of the three lepton flavors $\{e, \mu, \tau\}$. Figure \ref{fig:slepton} shows a representative production and decay diagram for this model scenario.

% \iffalse
%\subsection{Higgsino-Like LSP Scenario}
\iffalse
In the MSSM formalism, if the neutralinos and charginos are purely higgsino, then the mass splittings between them are of $\mathcal{O}(\mathrm{MeV})$. By introducing additional mixing with wino or bino states, these mass differences increase to $\mathcal{O}({10-}100\mathrm{GeV})$. Therefore, the production cross section in the higgsino scenario increases with the mass differences. Another important difference between the wino-bino and higgsino models is the final state kinematics. In the Wino-Bino case, the mass eigenvalue product $m(\neut) \cdot m(\neuo)$ can be either positive or negative. However, in the higgsino case, this product can only take negative values. In our simplified approach, the neutralino and chargino mixing matrices are fixed such that the neutralinos and charginos are higgsino states, regardless of the $\Delta m$ value. Furthermore, $\neuo$ is the LSP and $\ch$ is the NLSP. The $\ch$ mass is calculated as:
\begin{align}
    m(\ch) &= \frac{\neuo + \neut}{2}
\end{align}
Similar to the Wino-Bino $W^*/Z^*$ decays, the sleptons are much heavier than the neutralinos and charginos. Therefore, the decays proceed through virtual $W$ and $Z$ bosons such that the branching fraction ratios are unity:
\begin{align}
    \mathcal{B}(\ch \to \neuo W^{\pm,*} ) &= 1\\
    \mathcal{B}(\neut \to \neuo Z^{*} ) &= 1
\end{align}
\fi

\begin{figure}
    \centering
    \includegraphics[width=\linewidth]{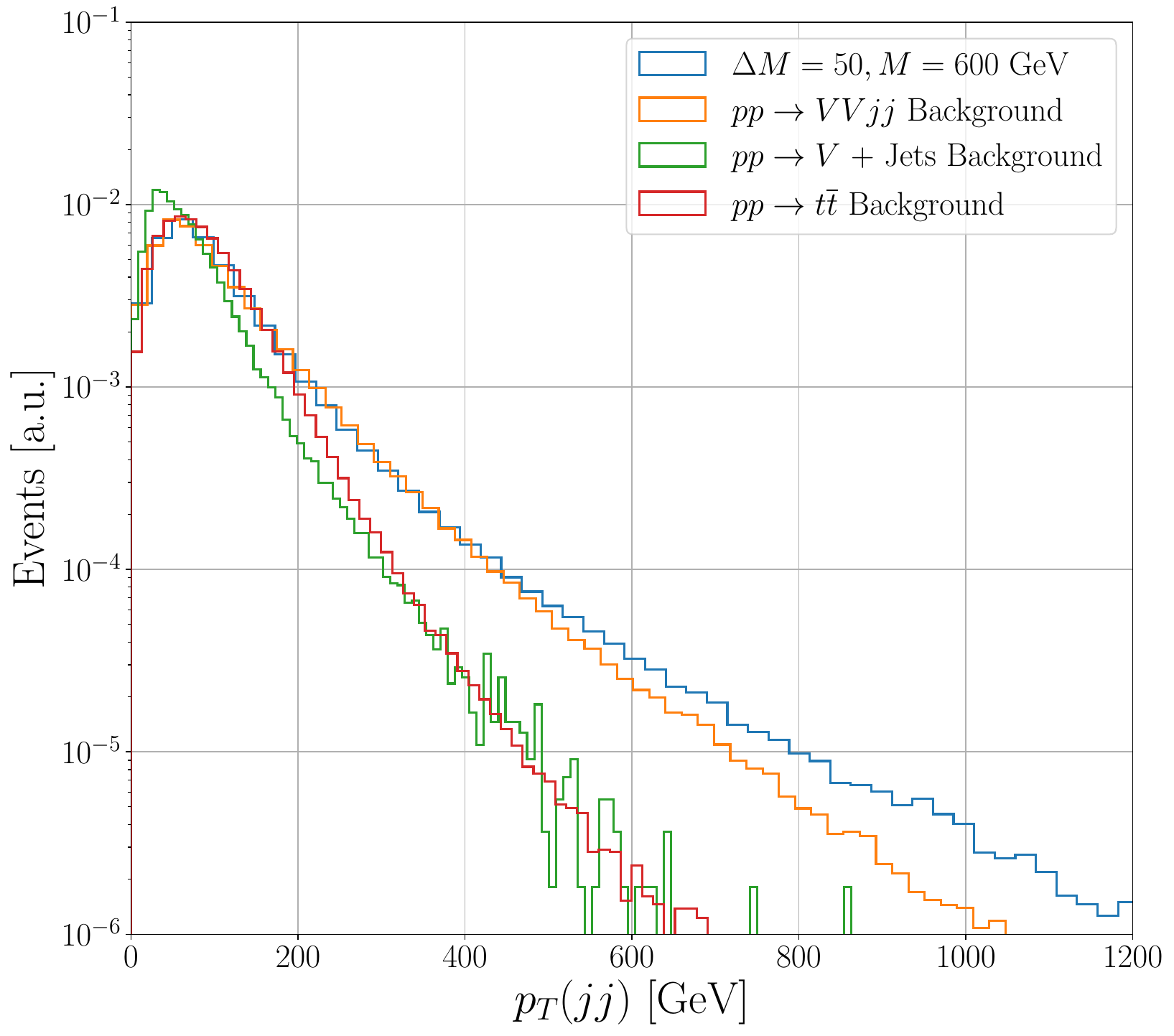}
    \caption{Dijet transverse momentum distributions for a $m(\ch)=600$ GeV and $\Delta m = 50$ Wino-Bino $W^*/Z^*$ signal and dominant SM backgrounds.}
    \label{fig:mjj}
\end{figure}

\begin{figure}
    \centering
    \includegraphics[width=\linewidth]{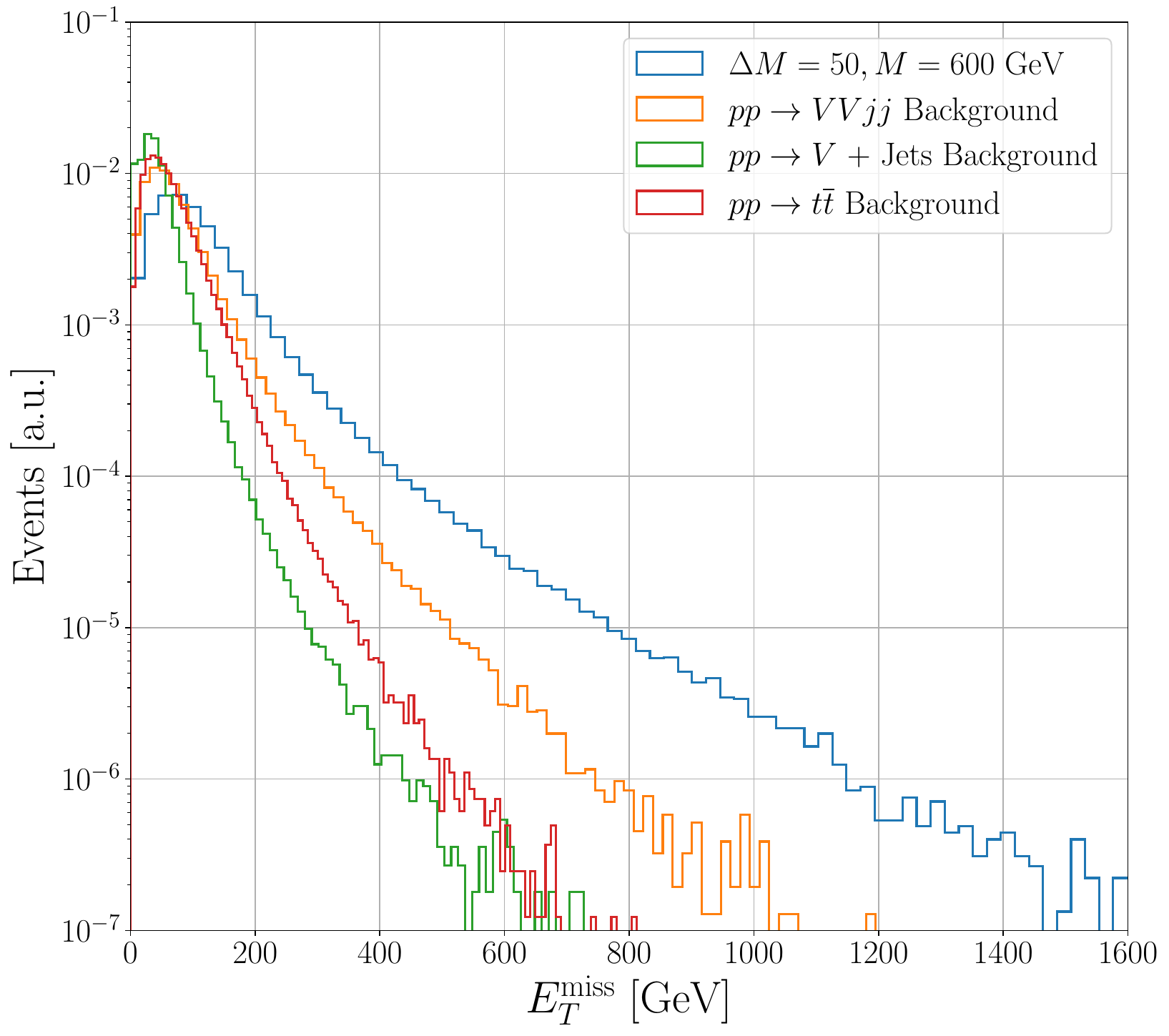}
    \caption{Missing transverse energy distributions for a $m(\ch)=600$ GeV and $\Delta m = 50$ Wino-Bino $W^*/Z^*$ signal and dominant SM backgrounds.}
    \label{fig:MET}
\end{figure}

\begin{figure}
    \centering
    \includegraphics[width=\linewidth]{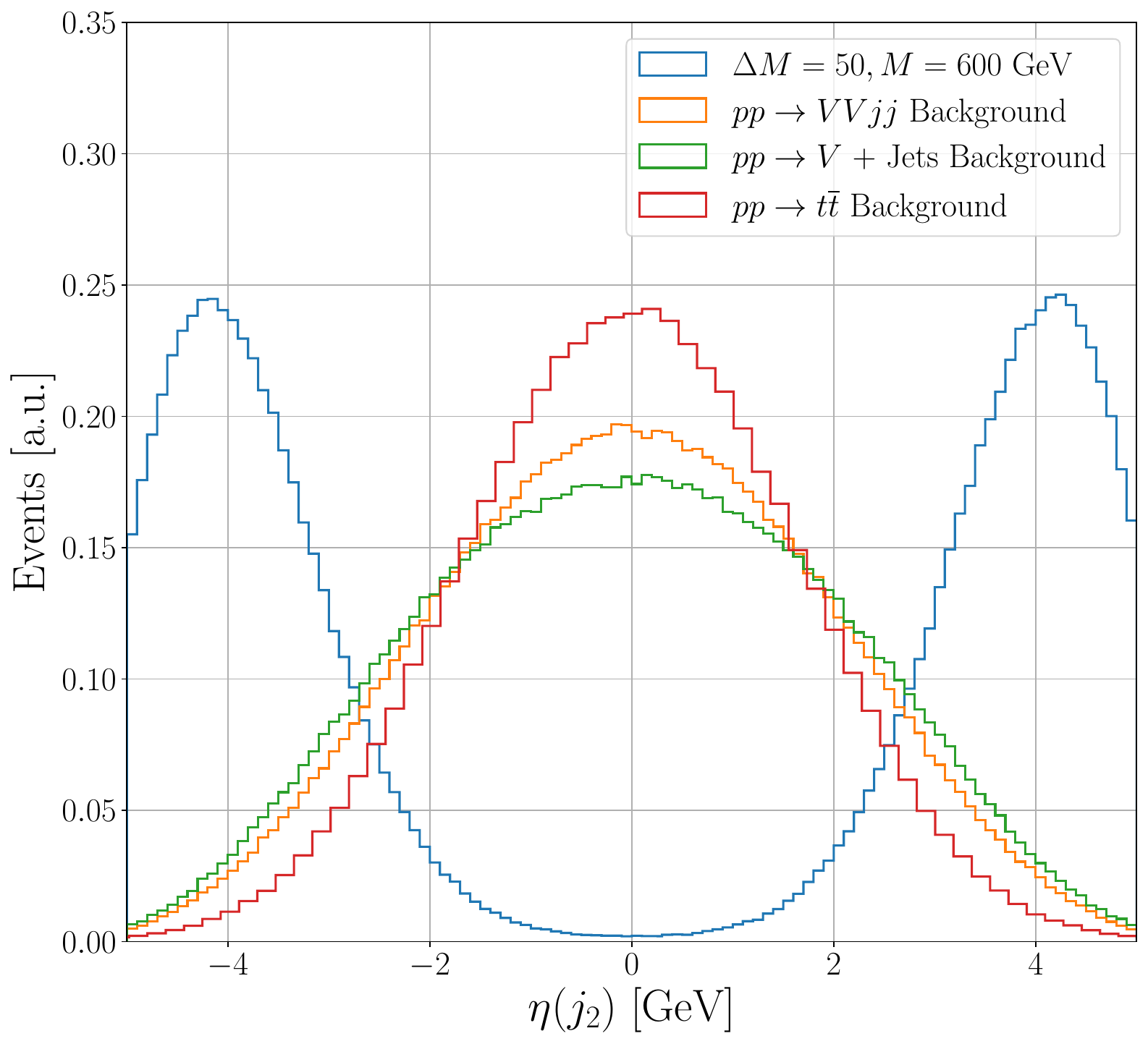}
    \caption{Pseudorapidity distributions of the jet with the second-highest transverse momentum for a $m(\ch)=600$ GeV and $\Delta m = 50$ Wino-Bino $W^*/Z^*$  signal and dominant SM backgrounds.}
    \label{fig:etaj2}
\end{figure}

\begin{figure}
    \centering
    \includegraphics[width=\linewidth]{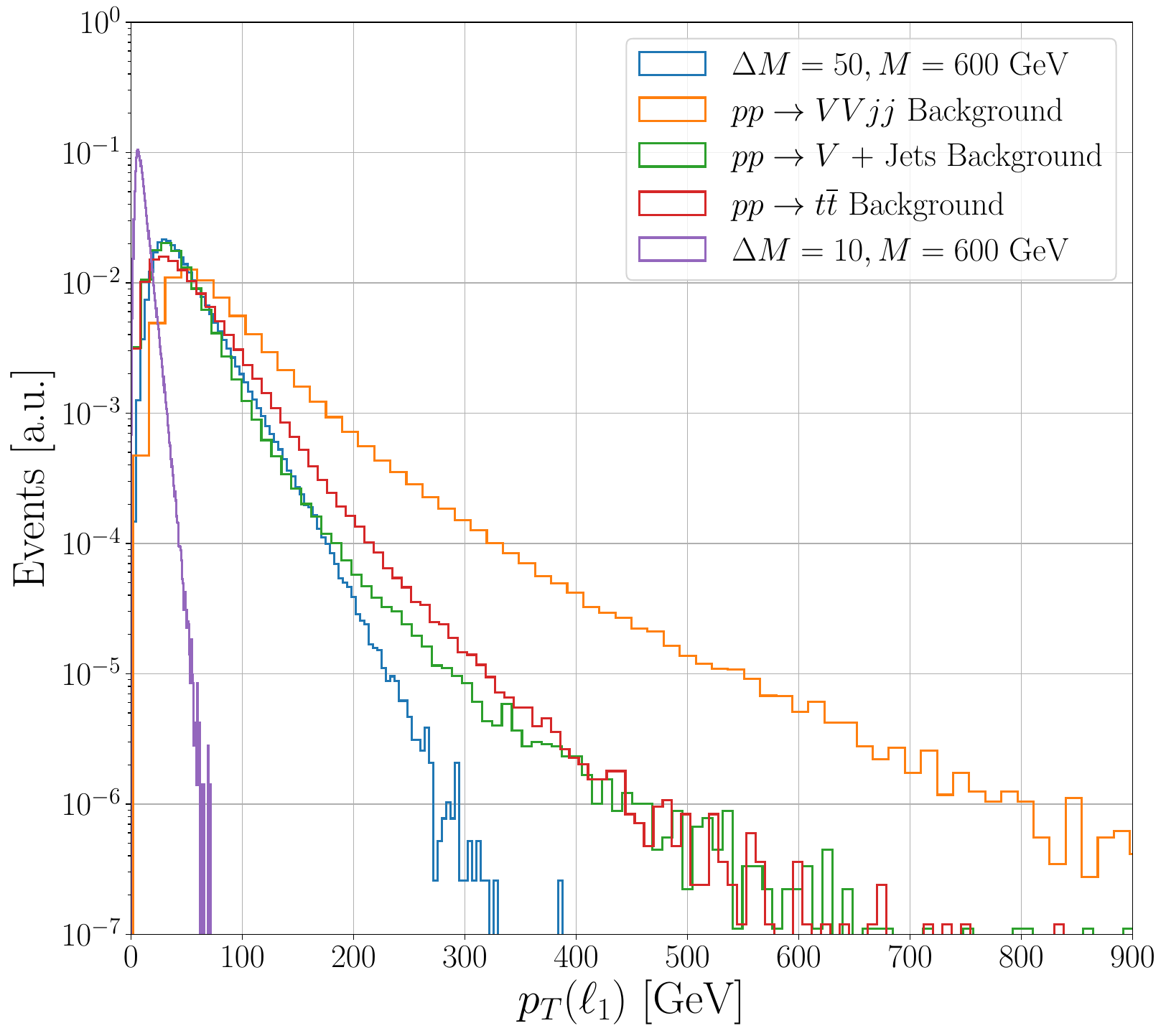}
    \caption{Transverse momentum distributions of the leading $p_T$-ordered lepton for $m(\ch)=600$ GeV and $\Delta m = 10, 50$ Wino-Bino $W^*/Z^*$ signals and dominant SM backgrounds.}
    \label{fig:pTl1}
\end{figure}

\begin{table}[h]
    \centering
    \begin{tabular}{c c}
    \hline
    \multicolumn{2}{c}{\textbf{Intitial Jet Selections}}  \\
    \hline
        $p_T(j)$ & $>30$ GeV \\
        $\abs{\eta(j)}$  & $<5.0$\\
        $\Delta R (j_1, j_2)$  & $>0.3$\\
    \hline 
    \multicolumn{2}{c}{\textbf{Topological Selections}} \\ \hline 
        $\abs{\eta(b)}$  & $<2.5$\\
         $p_T(b)$  & $>30$ GeV\\
         $N(b)$ & $=0$\\
         $\abs{\tau_h}$  & $<2.5$\\
         $p_T(\tau_h)$  & $>20$ GeV\\
         $N(\tau_h)$ & $=0$\\
        \hline
    \multicolumn{2}{c}{\textbf{Vector Boson Fusion Selections}} \\ \hline 
        $N(j)$ & $\ge 2$\\
        $\eta (j_1) \cdot \eta(j_2)$  & $< 0$\\
         $\abs{\Delta \eta (j_1, j_2)}$  & $>2.5$\\
         ${m(j_1, j_2)}$  & $>200$ GeV\\
        \hline
    \multicolumn{2}{c}{\textbf{1-Lepton Channel}} \\ \hline 
        $N(\ell)$ & $= 1$\\
        $\abs{\eta (\ell)}$  & $< 2.4$\\
         $p_T(\ell)$  & $>5$ GeV\\
        \hline
    \multicolumn{2}{c}{\textbf{2-Lepton Channel}} \\ \hline 
        $N(\ell)$ & $= 2$\\
        $\abs{\eta (\ell)}$  & $< 2.4$\\
         $p_T(\ell)$  & $>5$ GeV\\
    \hline
    \multicolumn{2}{c}{\textbf{3-Lepton Channel}} \\ \hline 
        $N(\ell)$ & $= 3$\\
        $\abs{\eta (\ell)}$  & $< 2.4$\\
         $p_T(\ell)$  & $>5$ GeV\\
    \hline
    \end{tabular}    
    \caption{Event selection criteria to maximize discovery potential at the LHC for the 1-lepton, 2-lepton, and 3-lepton channels.}
    \label{tab:eventsel}
\end{table}

\begin{table}[h]\label{tab:sel-eff}
\centering
\begin{tabular}{c |c c c }
\hline
 \boldmath $\Delta m$ \textbf{[GeV]}& \textbf{~1-Lepton~} & \textbf{~2-Lepton~} & \textbf{~3-Lepton~} \\ \hline
0.5 & 0.0342 & $\ll 1$\% & $\ll 1$\%     \\
2.5 & 0.1190 & 0.0318 &  $\ll 1$\%     \\
5.0 & 0.2781 & 0.0846 & 0.0092 \\
10 & 0.4081 & 0.3343 & 0.0835 \\
25 & 0.1134 & 0.4549 & 0.4223 \\
50 & 0.0293 & 0.3504 & 0.6196 \\
75 & 0.0135 & 0.3149 & 0.6714 \\ \hline
\end{tabular}
\caption{1, 2, and 3-lepton channel selection efficiencies as function of $\Delta m$ for a benchmark Wino-Bino $m(\neut) = m(\ch) = 600$ GeV scenario.}
\end{table}

In the studies presented in this paper, we consider final states containing two VBF-tagged jets, missing transverse momentum, and either one, two, or three light leptons (electrons or muons). Therefore, we consider the following sources of SM background: (i) production of ${t}$ quark pairs (${t}\overline{{t}}$); (ii) single Higgs boson $({h})$ production (gluon-gluon fusion, VBF, and associated production); (iii) $V=\{Z,W\}$ boson with associated jets; (iv) pure QCD multijet production; (v) diboson $VV=\{WW, ZZ, WZ\}$; and (vi) triboson events (${VVV}$). The three important SM backgrounds are ${t}\overline{{t}}$, $V$ + jets, and $VV$ with associated jets. The rest of the aforementioned background processes do not contribute meaningfully in our context, accounting for $< 1\%$ of the total background. 

The identification of leptons, light-quark or gluon jets, and bottom quarks is crucial for both identifying signal events and reducing SM background rates, thereby enhancing the discovery potential at the high-luminosity LHC. However, lepton reconstruction and identification may be challenging at the HL-LHC due to significant pileup. The impact of pileup on new physics discovery, particularly for VBF processes, and the importance of pileup mitigation at CMS and ATLAS, has been discussed in Ref.~\cite{CMS-PAS-FTR-13-014}. While the expected performance of the upgraded ATLAS and CMS detectors for the HL-LHC is beyond the scope of this work, reasonable expectations are provided by conservatively assuming some degradation in lepton and hadron identification efficiencies, using Ref.~\cite{CMS-PAS-FTR-13-014} as a benchmark, and considering the case of 140 average pileup interactions. 

The tracking efficiency for charged hadrons, which impacts both the jet clustering algorithm and the calculation of missing transverse momentum, is 97\% for $1.5 < |\eta| < 2.5$, decreasing to approximately 85\% at $|\eta| = 2.5$. For light leptons with $p_{T} > 5$ GeV and $|\eta|< 1.5$, the assumed identification efficiency is 95\% with a misidentification rate of 0.3\%~\cite{CMS-PAS-FTR-13-014,CMS_MUON_17001}. The performance degrades linearly with $\eta$ for $1.5 < |\eta| < 2.5$, with an identification efficiency of 65\% and a misidentification rate of 0.5\% at $|\eta| = 2.5$. These inefficiencies, arising from secondary $pp$ interactions, affect the accuracy of reconstructed lepton kinematics. Following Ref.~\cite{CMSbtag}, we consider the ``Loose'' working point of the DeepCSV algorithm~\cite{Bols_2020}, which gives a 70-80\% b-tagging efficiency and 10\% light quark mis-identification rate. The choice of the b-tagging working point is determined through an optimization process which maximizes discovery reach. 
 
% \begin{figure}[]
% \centering
% \includegraphics[width=0.45\textwidth]{figures/bg_Z_full.pdf}
% \caption{Representative Feynman diagram for a background event. A $Z$ boson is produced in association with a top quark through the fusion of a top, anti top pair from incoming protons. The $Z$ boson subsequently decays to a pair of muons and the two spectator top quarks decay semi-leptonically and purely hadronically to muons, neutrinos and jets, resulting in the same final states as the signal event.\label{fig:v}}
% \end{figure}

% \begin{figure}[]
% \centering
% \includegraphics[width=0.45\textwidth]{figures/M_mu_1_2.pdf}
% \caption{Invariant mass distribution of the muon pair with the highest and second highest transverse momentum.\label{fig:m_mu12}}
% \end{figure}

% \begin{figure}[]
% \centering
% \includegraphics[width=0.45\textwidth]{figures/PT_b1.pdf}
% \caption{Transverse momentum distribution of the leading $b$-quark jet candidate.\label{fig:pTb1}}
% \end{figure}

\section{Machine Learning Workflow}\label{sec:ML}
The analysis of signal and background events is done utilizing machine learning event classifiers. A machine learning-based approach offers sizeable advantages when compared to traditional event classification techniques.

Unlike conventional methods, machine learning models have the capability to simultaneously consider all kinematic variables, allowing them to efficiently navigate the complex and high-dimensional space of event kinematics. Consequently, machine learning models can effectively enact sophisticated selection criteria that take into account the entirety of this high-dimensional space. This makes them ideal for high-energy physics applications. 

For this study, we use a model inspired by the state-of-the-art TabNet \cite{Arik_Pfister_2021}
architecture. TabNet is a deep learning model specifically designed for tabular data, which combines the strengths of decision trees and neural networks by leveraging sequential attention to process features. Unlike traditional models such as Boost Decision Trees (BDTs) and Multi-Layer Perceptrons (MLPs), TabNet employs a self-supervised learning mechanism where it iteratively selects relevant features while ignoring irrelevant ones through its trainable feature mask. This adaptive feature selection enhances interpretability, which is particularly beneficial for high-dimensional datasets common in high-energy physics. Moreover, TabNet's sequential attention mechanism enables it to capture complex feature interactions without requiring extensive feature engineering. Unlike BDTs, TabNet also avoids relying heavily on gradient boosting and ensemble methods. In addition to outperforming traditional MLPs in terms of accuracy, TabNet also provides a more robust and interpretable model well-suited for the complex datasets often encountered in high-energy physics. 
The effectiveness of BDTs and MLPs has been validated in numerous experimental and phenomenological studies~\cite{Ai:2022qvs, ATLAS:2017fak, Biswas:2018snp,  Chung:2020ysf, ttZprime, Chigusa:2022svv,  Florez2023, Arganda2024, flórez2024probinglightscalarsvectorlike}. On the canonical Higgs Boson dataset \cite{misc_higgs_280}, where the task is distinguishing between Higgs boson processes vs. background, MLPs outperform BDT
variants even with very large ensembles \cite{Baldi_2014}. TabNet in turn outperforms MLPs \cite{Arik_Pfister_2021} with more compact representations, thus motivating its use in our study. In our experimentation, we found that this architecture provides a significant improvement over BDTs, extending 95\% exclusion bounds by as much as 150 GeV with $\mathcal{L}_\mathrm{int}=3000~\mathrm{fb}^{-1}$ over BDTs.

\begin{figure*}
    \centering
    \begin{minipage}{.33\textwidth}
  \centering
  \includegraphics[width=\linewidth]{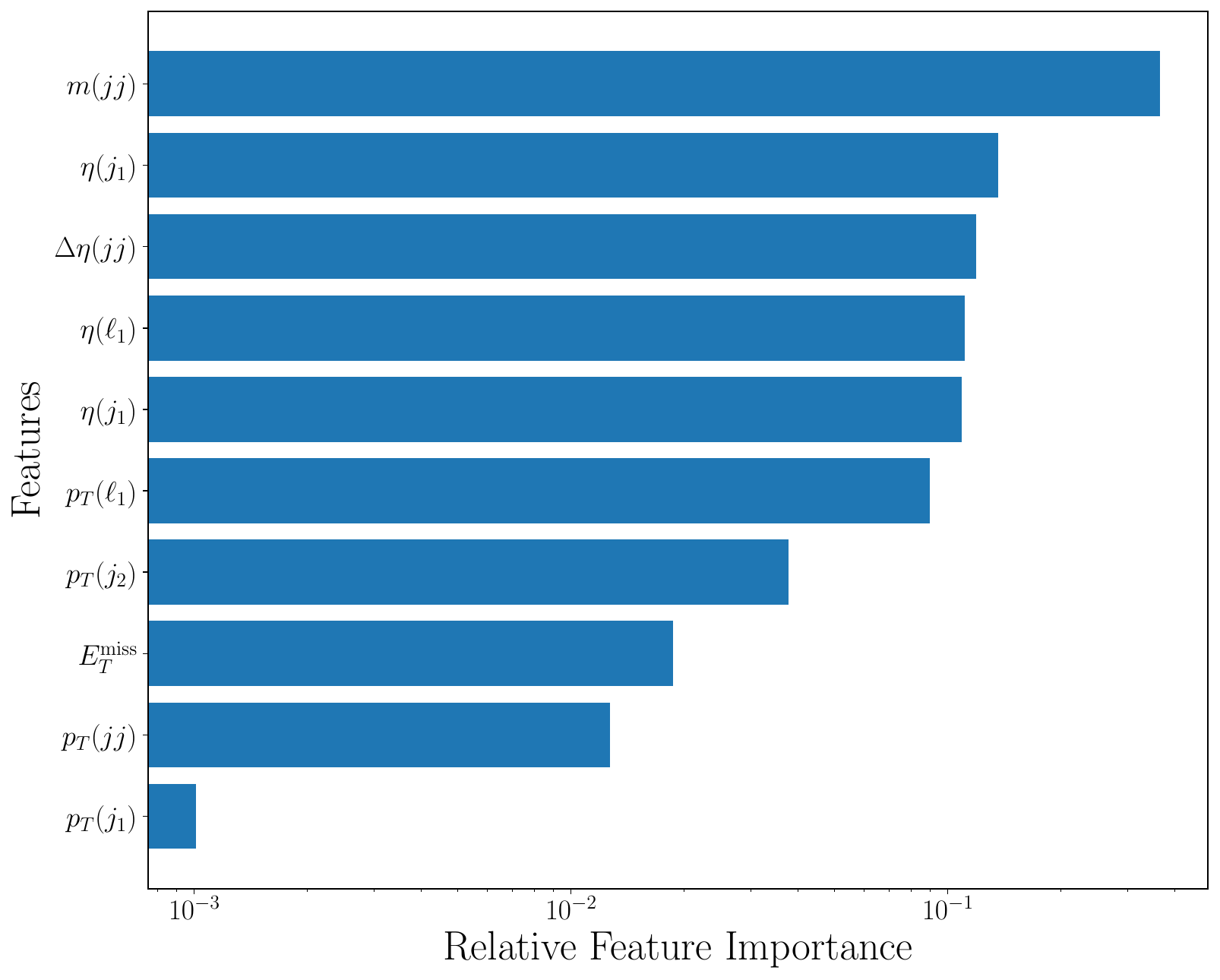}
\end{minipage}%
\begin{minipage}{.33\textwidth}
  \centering
  \includegraphics[width=\linewidth]{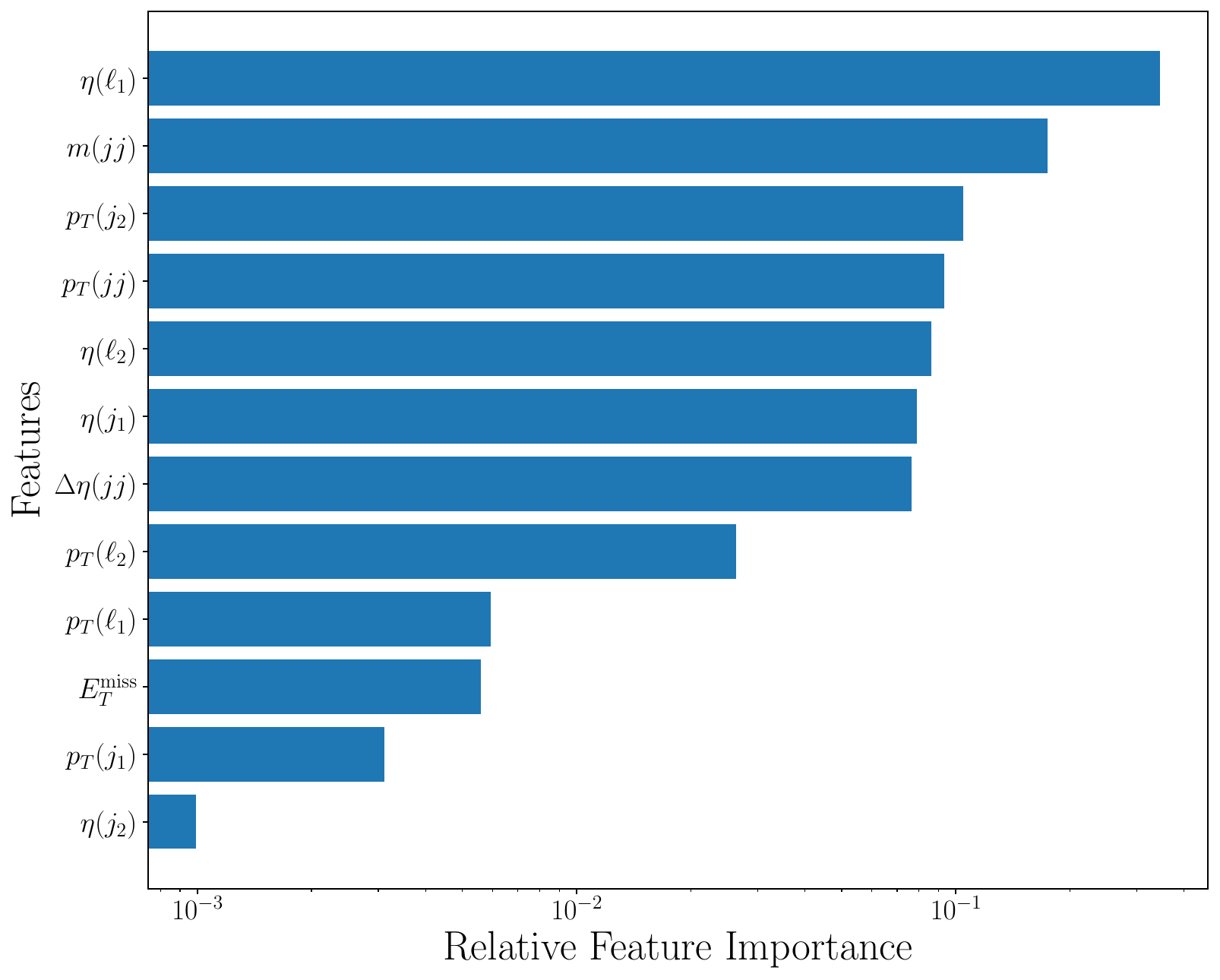}
\end{minipage}%
\begin{minipage}{.33\textwidth}
  \centering
  \includegraphics[width=\linewidth]{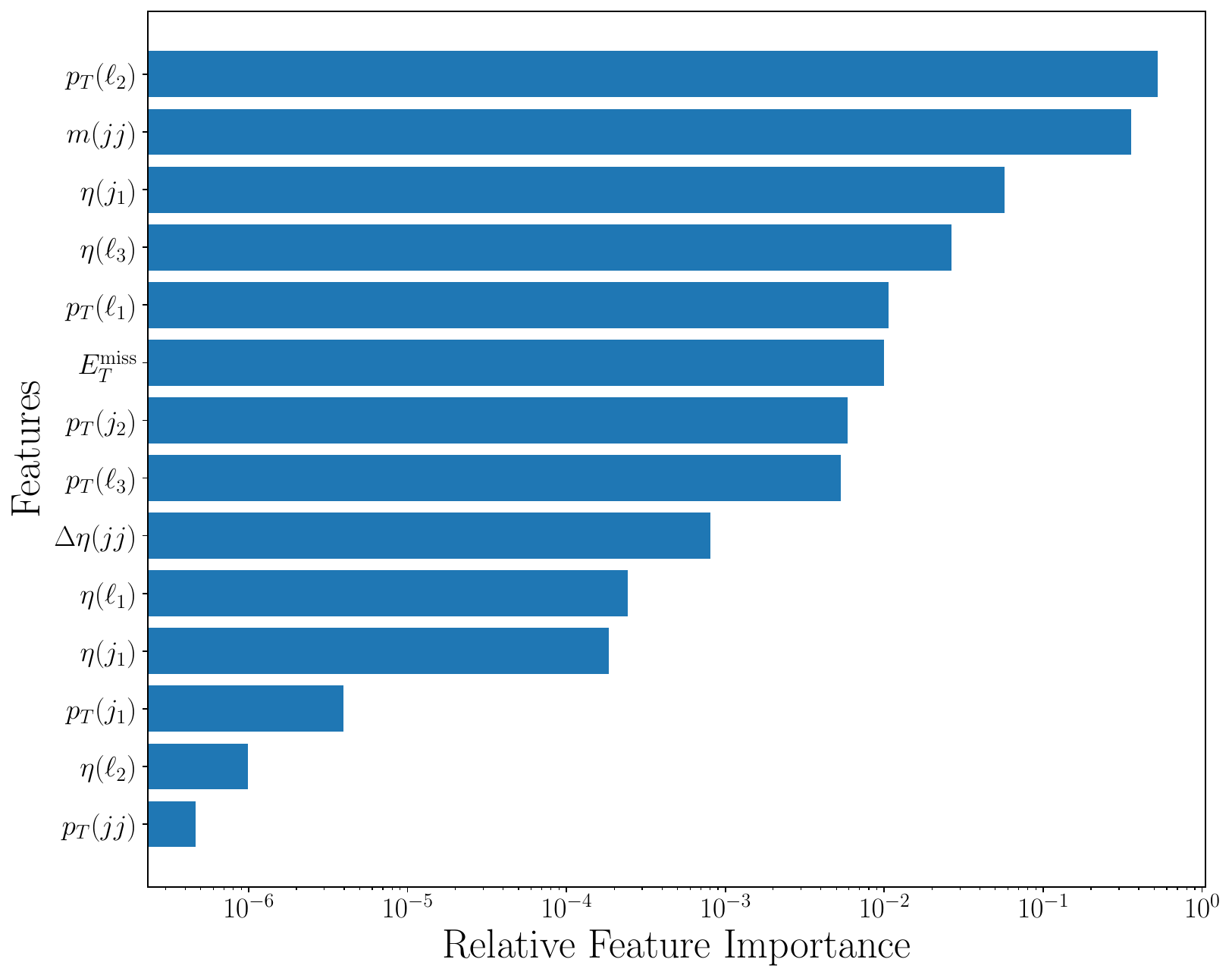}
\end{minipage}%
    \caption{Relative importance of features in training for a benchmark $\Delta m = 50$ and $m(\neuo)=m(\ch) =600 $ GeV Wino-Bino $W^*/Z^*$ scenario for the 1-lepton (left), 2-lepton (middle), and 3-lepton (right) channels respectively.}
    \label{fig:featimp}
\end{figure*}

For our machine learning workflow, kinematic variables such as particle and jet momenta, energies, and topologies are used as features to train the machine learning models. Figure \ref{fig:featimp} shows the particular features and their respective importance in the training process for a $\Delta m = 50$ and $m(\neuo)=m(\ch) =600 $ GeV benchmark in the Wino-Bino $W^*/Z^*$ scenario.

Training and evaluation of our ML models were carried out in a high-performance computing environment on an Nvidia A100 GPU. The canonical \texttt{PyTorch} \cite{paszke2019pytorchimperativestylehighperformance} deep learning framework was employed for implementing, configuring, training, and evaluating the models with a \texttt{scikit-learn} \cite{scikit-learn} wrapper for easy deployment. \texttt{PyTorch} and \texttt{scikit-learn} are well-regarded for their flexibility, ease-of-use, and performance in deep learning applications. For model hyperparameters, the following are found to be optimal. The size of the attention $n_a$ and decision $n_d$ layers is kept equal at $n_a=n_d = 8$. The coefficient $\gamma$ for feature reusage in the masks is set to $\gamma = 1.3$. The number of shared and independent gated linear units layers is kept at 2. The Adam \cite{kingma2017adammethodstochasticoptimization} optimizer with a learning rate $\alpha = 0.02$ is used in conjunction with the binary cross-entropy loss function. A train-test split of 90-10 is used with a batch size of $16384$ for training. The machine learning algorithm's output on the test set (i.e., data that is unseen during training) is used for the proceeding signal significance calculation. 

% \begin{table}[]
%     \centering
%     \begin{tabular}{c|c|c|c|c}
%     % \hline
%     \hline
%         Model & $\Delta m, m(\neut) $ & Scenario & Region &   AUC  \\
%         % \hline
%         \hline
%          \textbf{TabNet} & $50, 600$  & BW & $1\ell$ & \textbf{0.9989}\\
%          BDT & $50,600$ &  BW & $1\ell$ & 0.9800\\
%          \hline
%          \textbf{TabNet} & $50, 600$  & BW & $2\ell$ & \textbf{0.9989}\\
%          BDT & $50,600$ &  BW & $2\ell$ & 0.9875\\
%          \hline
%          \textbf{TabNet} & $50, 600$  & BW & $3\ell$ & \textbf{0.9988}\\
%          BDT & $50,600$ &  BW & $3\ell$ & 0.9987\\
%          \hline
%          \textbf{TabNet} & $10, 600$  & BW & $3\ell$ & \textbf{0.9999}\\
%          BDT & $10,600$ &  BW & $3\ell$ & 0.9998\\
%          \hline
%          % \hline

%     \end{tabular}
%     \caption{Area under the receiver operating characteristic (false positive rate v.s. true positive rate) curve (AUC) for several benchmark signal benchmark points. An AUC of 1 indicates perfect background rejection and signal acceptance. Entries in boldface indicate the better-performing model and AUC score.}
%     \label{tab:my_label}
% \end{table}

\section{Results}\label{sec:results}
Figures~\ref{fig:mjj},~\ref{fig:MET},~\ref{fig:etaj2}, and~\ref{fig:pTl1} show relevant kinematic distributions for the benchmark signal samples and the dominant SM backgrounds. Figure \ref{fig:mjj} shows the reconstructed dijet transverse momentum $p_T(jj)$ (normalized to unity) of the dominant SM backgrounds and a signal benchmark point corresponding to $m(\ch) = m(\neut) = 600$ GeV and $\Delta m = 50$ GeV. In rare cases ($\ll 1$\%) when there are more than two well-reconstructed and identified jet candidates with $p_{T} > 20$ GeV and $|\eta| < 5$, the dijet pair with the highest $m(jj)$ value is selected as the pair contributing to Figure~\ref{fig:mjj}. For the V+jets background, the jets are from soft initial state radiation and typically have average momentum near the kinematic reconstruction threshold, thus $p_{T}(jj)$ peaks at approximately 40 GeV. For the $t\bar{t}$ background, the jet candidates are $\textrm{b}$ quarks from top-quark decays (the b-quark jets fail the b-tagging identification criteria). Therefore, the average jet transverse momentum is expected to be $\frac{m_{\mathrm{t}} - m_{\mathrm{W}}}{2} \approx 45$ GeV, and thus the average $p_{T}(jj)$ is about 90 GeV. On the other hand, the signal distribution and VVjj distribution (from vector boson scattering) have a harder jet $p_{T}$ spectrum typical of VBF/VBS processes, but the jets are in opposite hemispheres of the detector. Therefore, the $p_{T}(jj)$ distributions have a longer tail representative of a harder jet $p_{T}$ spectrum, but there can be some cancellations when performing the vectorial sum of the jet momentum vectors, leading to a peak below 100 GeV. 

Similar to current SUSY searches at ATLAS and CMS, we infer the production of $\neuo$ candidates at the LHC indirectly through the measurement of momentum imbalance in the transverse plane of the detectors. The reconstructed $E_{T}^\mathrm{miss}$ is defined as the norm of the negative vectorial sum of the transverse momenta of visible particle candidates. 
%\begin{align}
%    E_{T}^\mathrm{miss} &= \norm{-\sum_{i\in \mathrm{visible}}p_{T,i} }
%\end{align}
Figure \ref{fig:MET} shows the $E_{T}^\mathrm{miss}$ distributions (normalized to unity) of the dominant SM backgrounds and a signal benchmark point corresponding to $m(\ch) = m(\neut) = 600$ GeV and $\Delta m = 50$ GeV. As expected, the SM backgrounds have neutrinos from W boson decays, which result in an average $E_{T}^\mathrm{miss}$ of $\approx m_{W}/2$, while the signal distribution shows a harder spectrum on average due to the heavy LSP mass. 

Meanwhile, Figure \ref{fig:pTl1} shows the leading lepton $p_T$ distribution for dominant SM backgrounds and signal benchmark points corresponding to $m(\ch) = m(\neut) = 600$ GeV and $\Delta m = 50, 10$ GeV. The leptons from signal processes have an average transverse momentum of $p_{T} \approx \Delta m/3$, while the SM backgrounds have an average lepton $p_{T}$ of $m_{W}/2$ or $m_{Z}/2$.

Finally, Figure \ref{fig:etaj2} shows the subleading jet pseudorapidity distributions. As expected, the SM backgrounds mainly feature jets with $\eta \approx 0$, which are central in the detector and form dijet pairs with small $|\Delta\eta(jj)|$. In contrast, the signal distribution is characterized by jets that are closer to the proton beam line and form dijet pairs with large $|\Delta\eta(jj)|$.

Variables such as the ones in Figures~\ref{fig:mjj}-\ref{fig:pTl1} are used as inputs to the TabNet machine learning model. For this purpose, our workflow involves the use of a specialized \texttt{MadAnalysis Expert Mode} C++ script~\cite{CONTE2013222} which extracts essential kinematic and topological information from the simulated samples. The script transforms relevant variables contained within these files and transforms them into a structured and informative CSV (Comma-Separated Values) format that can be used to train our machine learning models. To properly account for the differential significance of various events, we apply cross-section weighting. This ensures that the relative importance of signal and background events is appropriately balanced in the dataset. The prepared and weighted datasets are then passed to our \texttt{MadAnalysis Expert Mode} C++ script, where the simulated signal and background events are initially filtered, before being passed to the CSV file for use by the machine learning algorithm. The filtering process requires that events be classified into three distinct search channels contingent on the lepton multiplicity and $p_T$. This decision is motivated by the fact lepton multiplicity and average $p_T$ depend strongly on $\Delta m$ as seen in Figure \ref{fig:pTl1}. Lower (higher) values of $\Delta m$ result in a softer (harder) lepton $p_T$ spectrum, and consequently a lower (larger) probability of reconstructing and identifying multiple leptons. The filtering criteria are summarized in Table \ref{tab:eventsel}. We require identified leptons to have $p_{T} > 5$ GeV and $|\eta| < 2.4$. An additional $\tau_{\textrm{h}}$ (b-jet) veto is applied by rejecting events containing a well-identified $\tau_{\textrm{h}}$ (b) candidate with $p_{T} > 20$ ($30$) GeV and $|\eta| < 2.5$. The VBF signal topology is characterized by the presence of two jets in the forward direction, in opposite hemispheres, and with large dijet invariant mass. On the other hand, the jets in background events are mostly central and have small dijet invariant masses. Therefore, the VBF filtering criteria is imposed by requiring at least two jets with $p_{T} > 30$ GeV and $|\eta| < 5$. All pairs of jet candidates passing the above requirements and having $|\Delta\eta(jj)| > 2.5$ and $\eta(j_{1} \times \eta(j_{2}) < 0$ are combined to form VBF dijet candidates, and selected dijet candidates are required to have $m(jj) > 200$ GeV. Some consideration has also been placed on the existence of suitable experimental triggers. Although a comprehensive study to address suitable and available HL-LHC triggers is beyond the scope of this work, we note that the CMS non-VBF compressed SUSY search in Ref. \cite{CMS:2018kag} utilized a soft lepton plus  $E^{\mathrm{miss}}_T$ trigger. As such, we deem it reasonable that an HL-LHC trigger that utilizes the VBF topology (in conjunction with a soft lepton and/or $E^{\mathrm{miss}}_T$) be developed for the proposed selections in this paper.

Table~\ref{tab:sel-eff} shows the fraction of signal events which satisfy the filtering criteria, as a function of $\Delta m$, for a $\tilde{\ch}$ mass of 600 GeV in the Wino-Bino WZ scenario. We refer to this filtering criteria as pre-selections. The efficiency of the 1-lepton (2-lepton) pre-selections is about 3.4 ($\ll 1$)\% for $\Delta m = 0.5$ GeV, increases to 40.8 (33.4)\% at $\Delta m = 10$ GeV, decreases (increases) to 11.3 (45.5)\% at an intermediate $\Delta m$ value of 25 GeV, and drops back down to 1.35 (31.5)\% in the highest value of $\Delta m$ considered in this paper. On the other hand, the 3-lepton pre-selection efficiency is $< 1$\% for $\Delta m < 5 GeV$, and increases to about 62\% at $\Delta m = 50 GeV$. Table~\ref{tab:sel-eff} highlights why the TabNet training and testing is separated into three distinct search channels contingent on the lepton multiplicity. The 1-lepton channel is expected to be the most relevant channel at low $\Delta m$, the 2-lepton channel is expected to be the most important channel at intermediate values of $\Delta m$, and the 3-lepton channel in the higher mass gap scenarios.

In addition to these aforementioned variables in Figures~\ref{fig:mjj}-\ref{fig:pTl1}, 
several other kinematic variables were included as inputs to the TabNet algorithm. In particular, 10-14 such variables were used in total, depending on the lepton multiplicity, 
and these included the momenta of jets and light lepton candidates; the missing transverse momentum; the largest invariant mass of pairs of jets in each event; the dijet $p_{T}(jj)$ of the jet pair with the largest $m(jj)$; and angular difference $|\Delta\eta(jj)|$ between the jet pair with the largest dijet mass in each event. 
%and $\mathrm{b}$ jets; and transverse masses derived from the lepton-$p^{miss}_{\mathrm{T}}$ pair and lepton-$p^{miss}_{\mathrm{T}}$-$\mathrm{b}$ triplets. 
As mentioned above, the variables $m(jj)$, $|\Delta\eta(jj)|$, $p_{T}(jj)$, and $p_{T}(j)$ are the distinguishing characteristics of the VBF process, while the lepton kinematics and the missing transverse momentum target small $\Delta m$ and the LSP mass, respectively. The TabNet learning model returns the discriminating power of each of its inputs, and the feature importance for each variable is shown in Figure~\ref{fig:featimp} for a signal benchmark point with $\Delta m = 50$ and $m(\neuo)=m(\ch) =600 $ GeV in the Wino-Bino $W^*/Z^*$ scenario.

\begin{figure*}
    \centering
    \begin{minipage}{.33\textwidth}
  \centering
  \includegraphics[width=\linewidth]{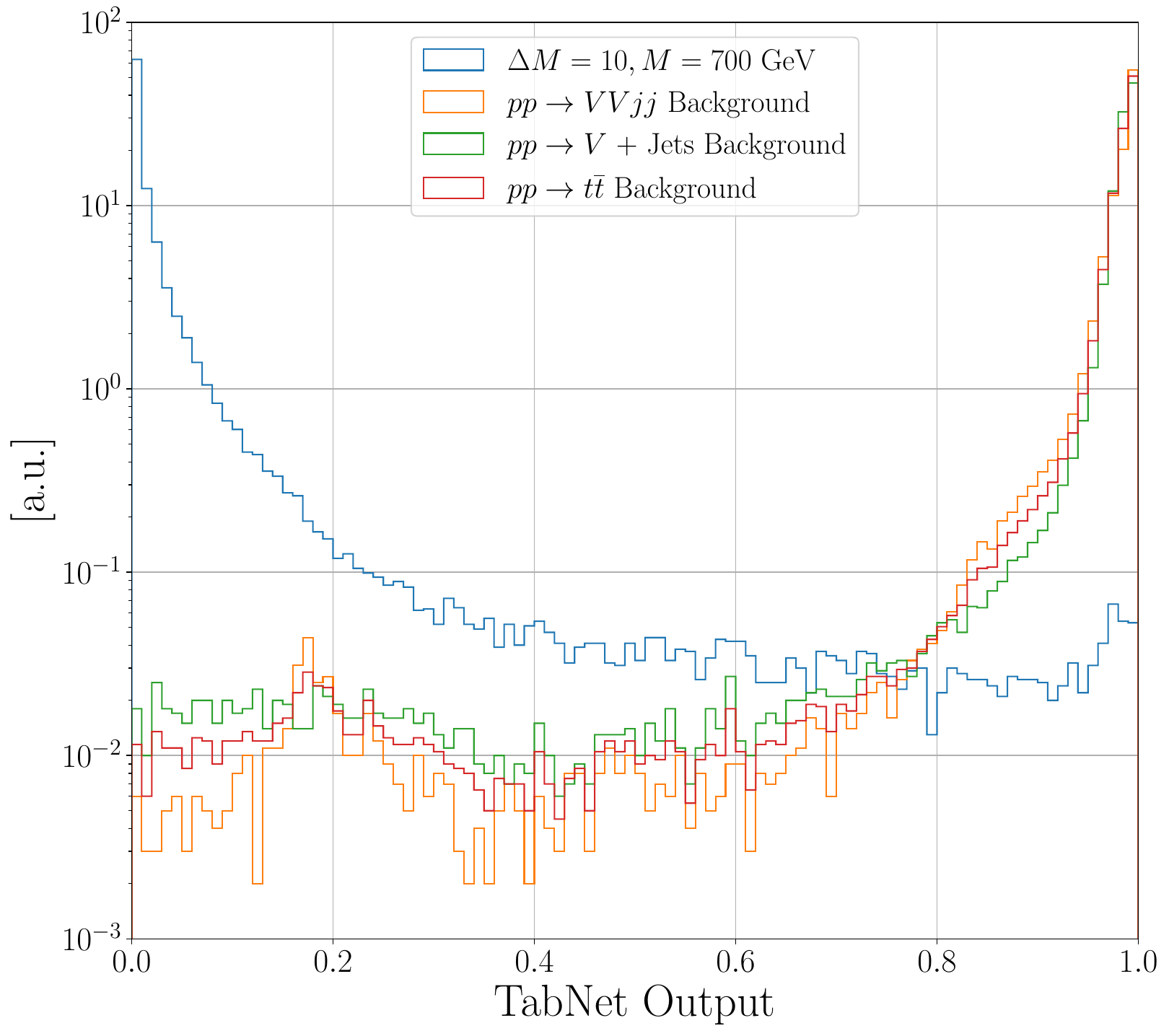}
\end{minipage}%
\begin{minipage}{.33\textwidth}
  \centering
  \includegraphics[width=\linewidth]{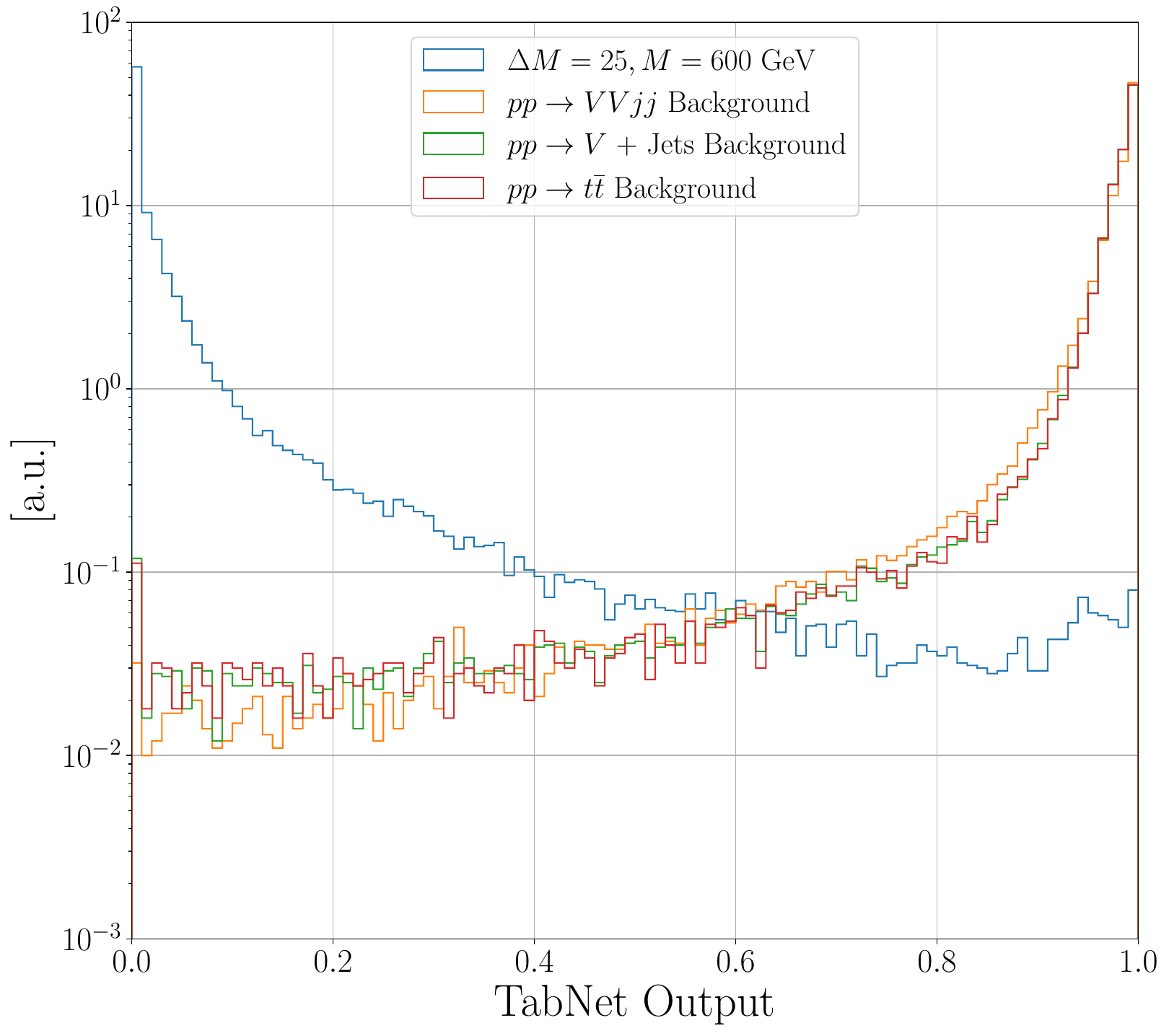}
\end{minipage}%
\begin{minipage}{.33\textwidth}
  \centering
  \includegraphics[width=\linewidth]{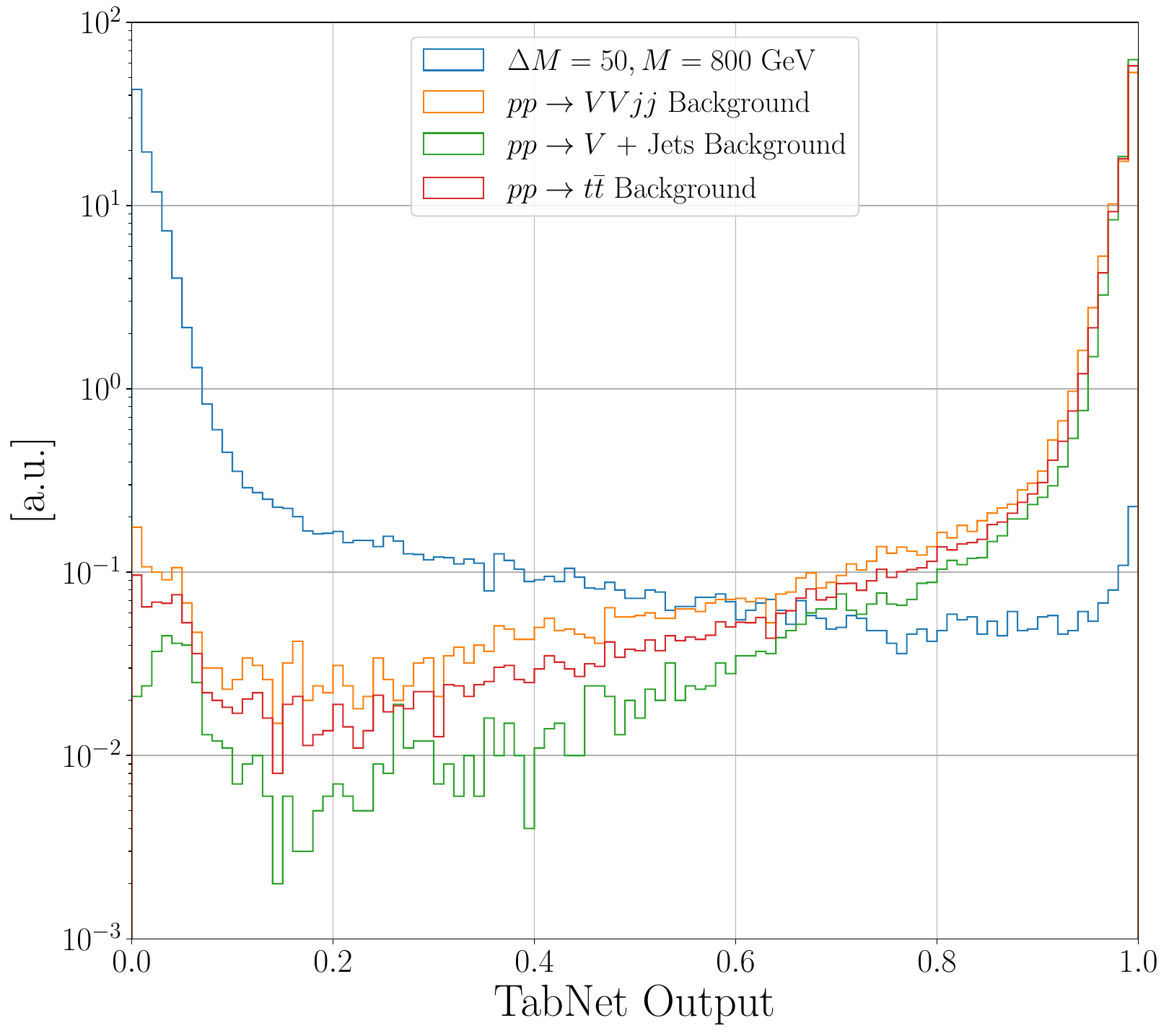}
\end{minipage}%
    \caption{TabNet output distributions for benchmark Wino-Bino $W^*/Z^*$ signal scenarios and dominant backgrounds for the 1 (left), 2 (middle), and 3-lepton (right) channels. The distributions are normalized to unity.}
    \label{fig:tabnetout}
\end{figure*}

% \begin{figure}
%     \centering
%     \includegraphics[width=\linewidth]{figures/2lepton_TabNet.pdf}
%     \caption{TabNet output distribution for a benchmark $\Delta m = 25$ and $m(\neuo)=m(\ch) =600 $ GeV Wino-Bino $W^*/Z^*$ scenario and dominant backgrounds in the 2-lepton channel. The distributions are normalized to unity.}
%     \label{fig:tabnetout}
% \end{figure}

As noted previously, events passing the pre-selections described above are used as input for the machine learning algorithm, which classifies them as signal or background, using a probability factor. Figur \ref{tabnetout} displays the distributions of the TabNet algorithm's output for both a representative signal benchmark and the dominant SM backgrounds. The TabNet output ranges from 0 to 1, representing the likelihood of an event being either signal-like (output near 0) or background-like (output near 1). The figure allows the calculation of the true positive rate, which is the probability of correctly identifying signal events, and the false positive rate, the probability of incorrectly identifying SM background events. For instance, in the 2-lepton channel, selecting events with TabNet output $< 0.25$ (0.5) results in a 99.6 (99.1)\% probability of selecting signal events for the Wino-Bino $W^{*}/Z^{*}$ scenario with $m(\ch) = 600$ GeV and $\Delta m = 25$ GeV, and a background selection probability of 5.7 (2.4)\%.

The full spectrum (entire range from 0 to 1) of the outputs from the TabNet machine learning algorithm is used to perform a profile-binned likelihood fit to estimate the expected signal significance. We consider three different integrated luminosity scenarios with $\mathcal{L}_{\mathrm{int}}=137$, $300$, and 3000 $\mathrm{fb^{-1}}$ corresponding to current, end of Run 3, and the HL-LHC. For the signal significance calculations, the output TabNet histograms are normalized to $N = \mathcal{L}_\text{int}\cdot \sigma \cdot \epsilon$ where $\epsilon$ is a factor taking into account selection and reconstruction efficiencies. The significance is then calculated using the expected bin-by-bin yields of the TabNet output distribution in a profile likelihood fit~\cite{Butterworth:2015oua}. The expected signal significance $Z_\text{sig}$ is calculated using the probability of obtaining the same test statistic for the signal plus background and the signal-null hypotheses, defined as the local $p$-value. Then, the significance corresponds to the point where the integral of a Gaussian distribution between $Z_\text{sig}$ and $\infty$ results in a value equal to the local $p$-value.

The estimation of $Z_\text{sig}$ incorporates systematic uncertainties. The uncertainty values have been included as nuisance parameters, considering lognormal priors for normalization and Gaussian priors for uncertainties associated with the modeling of the TabNet output shapes. The systematic uncertainties that have been included result from experimental and theoretical constraints. A 1-5\% systematic uncertainty, depending on the simulated MC sample, has been included to account for the choice of Parton Distribution Function (PDF) set. The systematic uncertainty effect was measured following the PDF4LHC~\cite{Butterworth:2015oua} recommendations. This systematic uncertainty has a small impact on the expected event yields for signal and background, but it does not affect the shape of the TabNet output distribution significantly (effects on the shape are less than the bin-by-bin statistical uncertainties). The PDF uncertainties are treated as uncorrelated across signal and background processes, and correlated across TabNet output bins for a given process. We also accounted for theoretical uncertainties due to the lack of higher-order contributions to the signal cross sections, which could affect the pre-selection efficiencies and the shapes of kinematic variables used by the BDT algorithm. To estimate this uncertainty, we varied the renormalization and factorization scales by a factor of 2 and examined the resulting changes in the bin-by-bin yields of the TabNet distributions. The uncertainties were found to be no greater than 2--5\%, depending on $m(\ch)$, $\Delta m$, and the TabNet output bin.

Regarding experimental uncertainties, following experimental measurements from CMS on the estimation of the integrated luminosity, a conservative 3\% effect has been included \cite{lumiRef}, which is treated as fully correlated across processes and correlated across TabNet output bins. For light leptons, we include a 2\% uncertainty associated with reconstruction, identification, and isolation requirements, and a 3\% systematic uncertainty to account for scale and resolution effects on the momentum and energy estimation, independent of $p_{T}$ and $\eta$ of the leptons. These lepton uncertainties are treated as correlated across signal and backgrounds with genuine leptons, and correlated across TabNet output bins for a given process. We consider jet energy scale uncertainties ranging from 2-5\%, contingent on $\eta$ and $p_T$, resulting in shape-based uncertainties on the TabNet output distribution, varying from 1-3\% depending on the bin. The methodology for calibrating the SM backgrounds with actual pp data is outside the scope of this work. However, to account for this effect, an additional 10\% systematic uncertainty, independent of TabNet output bin is included to account for errors in the signal and background predictions. This uncertainty is treated as uncorrelated across signal and background processes. Considering all the various sources of systematic uncertainties our conservative estimate yields a total effect of about 20\%.

In addition to the above experimental uncertainties, we also take into account statistical uncertainties arising from the number of simulated signal and background events that pass the selections and fall into a specific bin of the TabNet output histogram. These uncertainties vary between 1\% and 35\%, depending on the process and the particular TabNet output bin. These uncertainties are uncorrelated across processes and uncorrelated across TabNet output bins.

\begin{figure*}
    \centering
    \begin{minipage}{.5\textwidth}
  \centering
  \includegraphics[width=\linewidth]{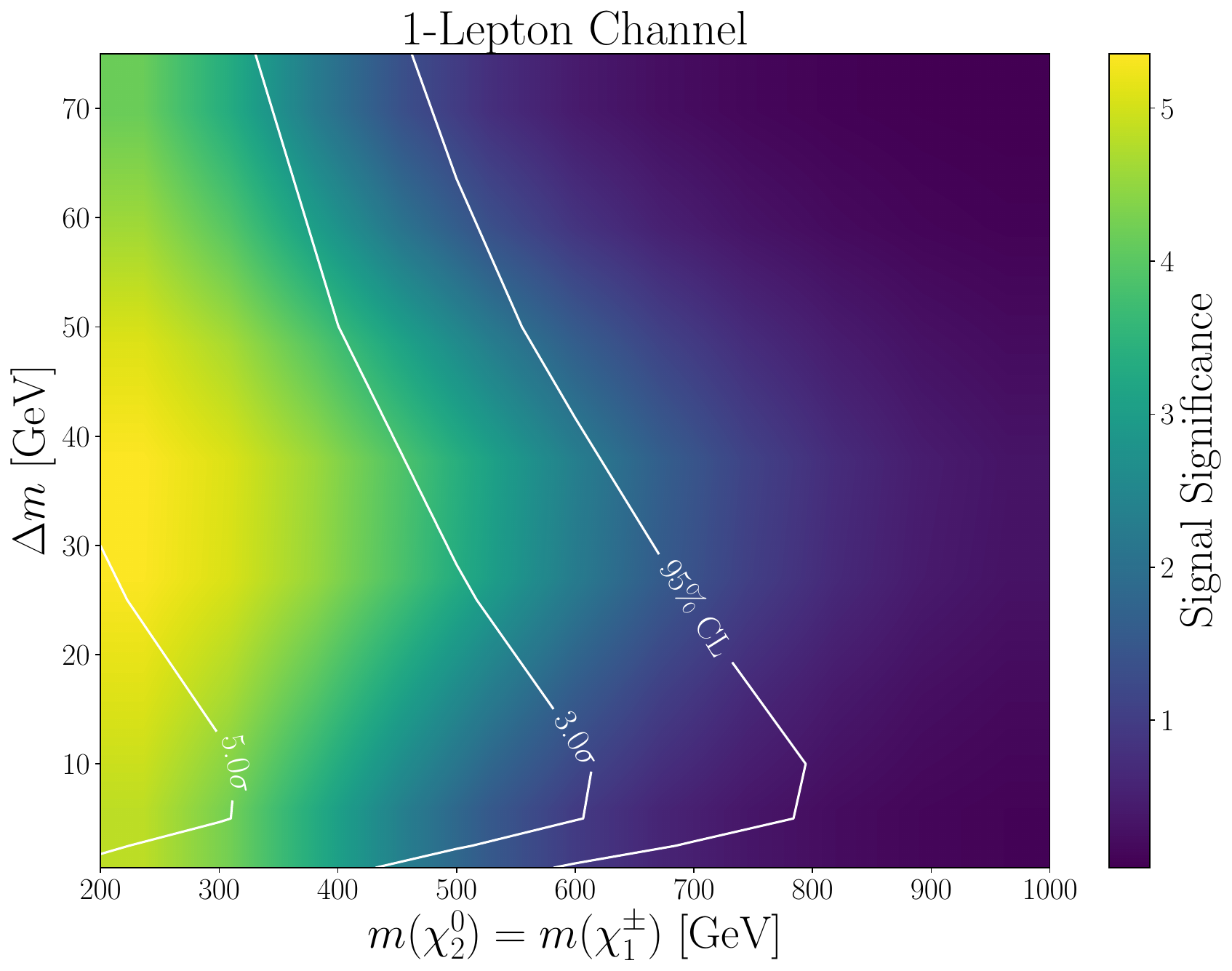}
\end{minipage}%
\begin{minipage}{.5\textwidth}
  \centering
  \includegraphics[width=\linewidth]{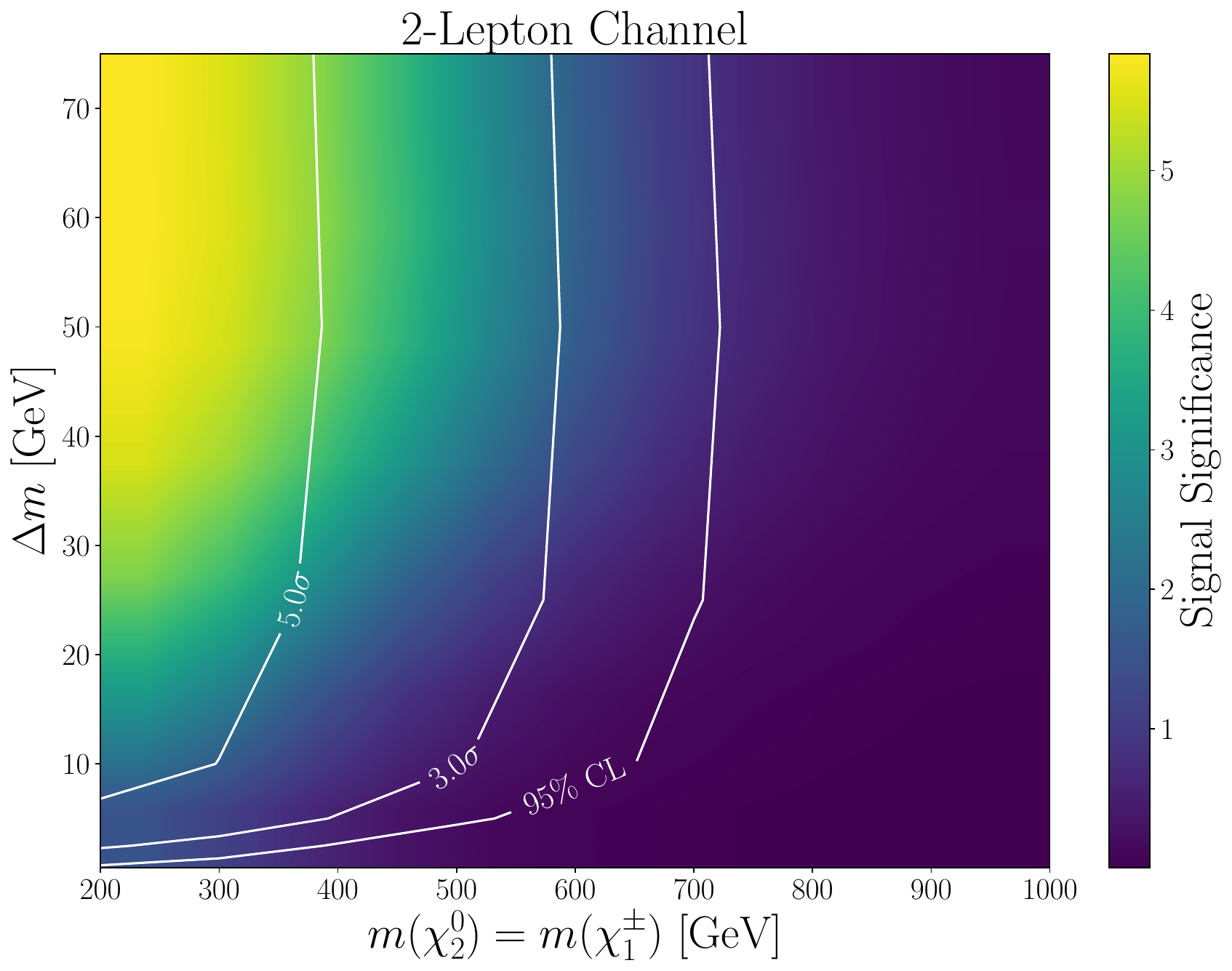}
\end{minipage}%

\begin{minipage}{.5\textwidth}
  \centering
  \includegraphics[width=\linewidth]{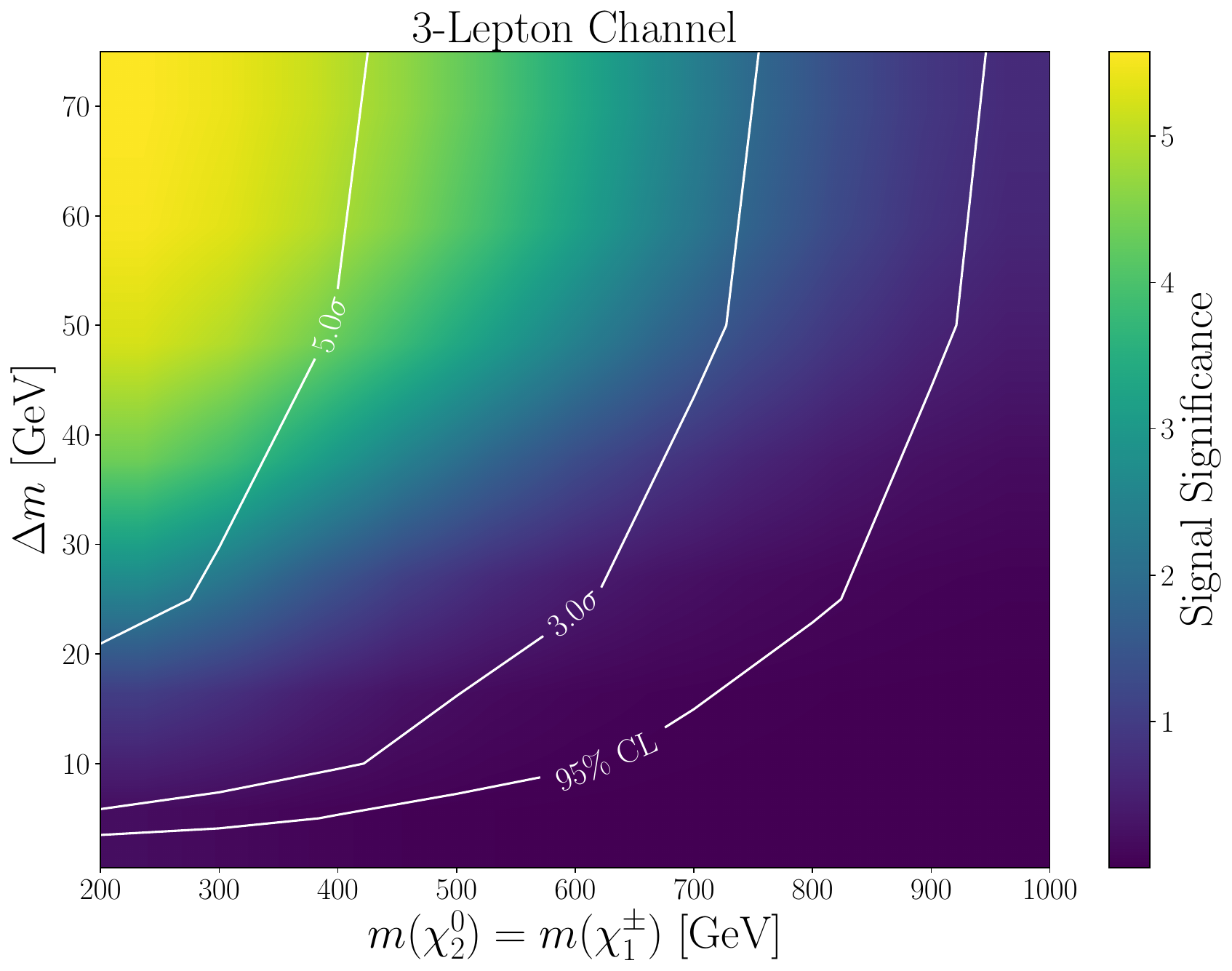}
\end{minipage}%
\begin{minipage}{.5\textwidth}
  \centering
  \includegraphics[width=\linewidth]{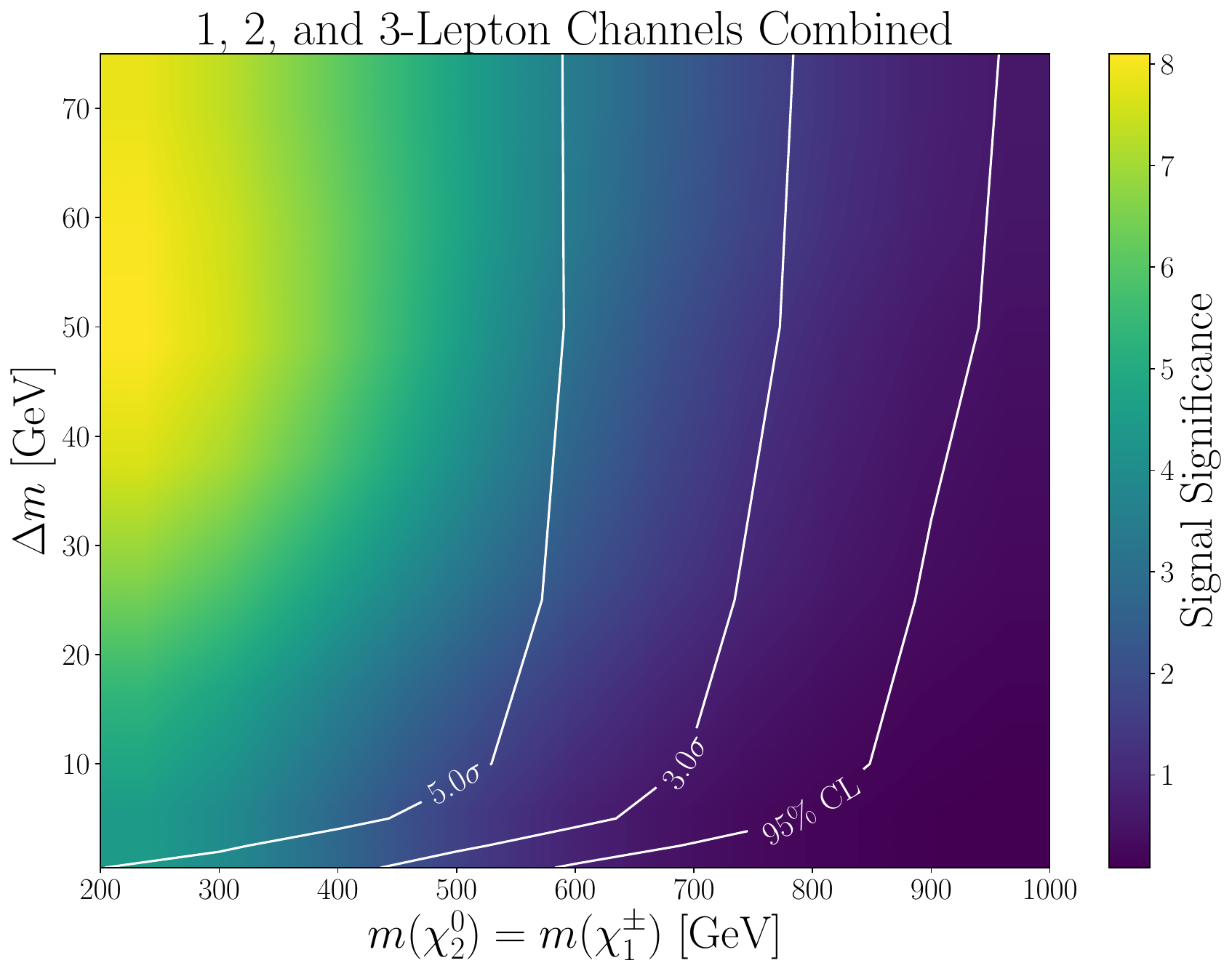}
\end{minipage}%
    \caption{Projected signal significance contours for the 1-lepton (top-left), 2-lepton (top-right), 3-lepton (bottom-left), and combined 1, 2, and 3-lepton (bottom-right) channels as a function of the $\ch$ mass and the mass difference $\Delta m$ for the Wino-Bino $W^*/Z^*$ scenario assuming an integrated luminosity of 3000 $\mathrm{fb}^{-1}$ corresponding to the HL-LHC.}
    \label{fig:SSBW3000}
\end{figure*}

\begin{figure*}
    \centering
    \begin{minipage}{.5\textwidth}
  \centering
  \includegraphics[width=\linewidth]{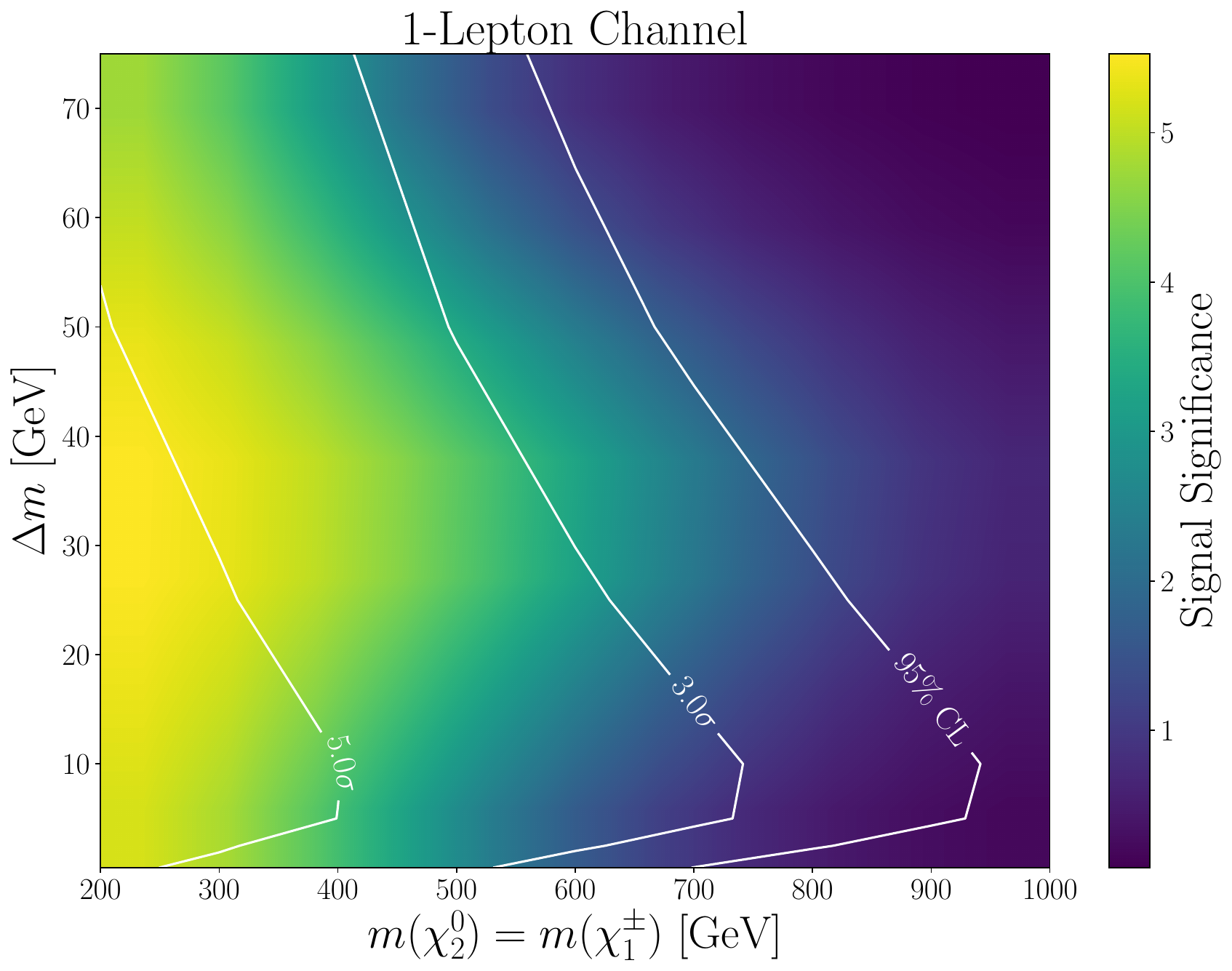}
\end{minipage}%
\begin{minipage}{.5\textwidth}
  \centering
  \includegraphics[width=\linewidth]{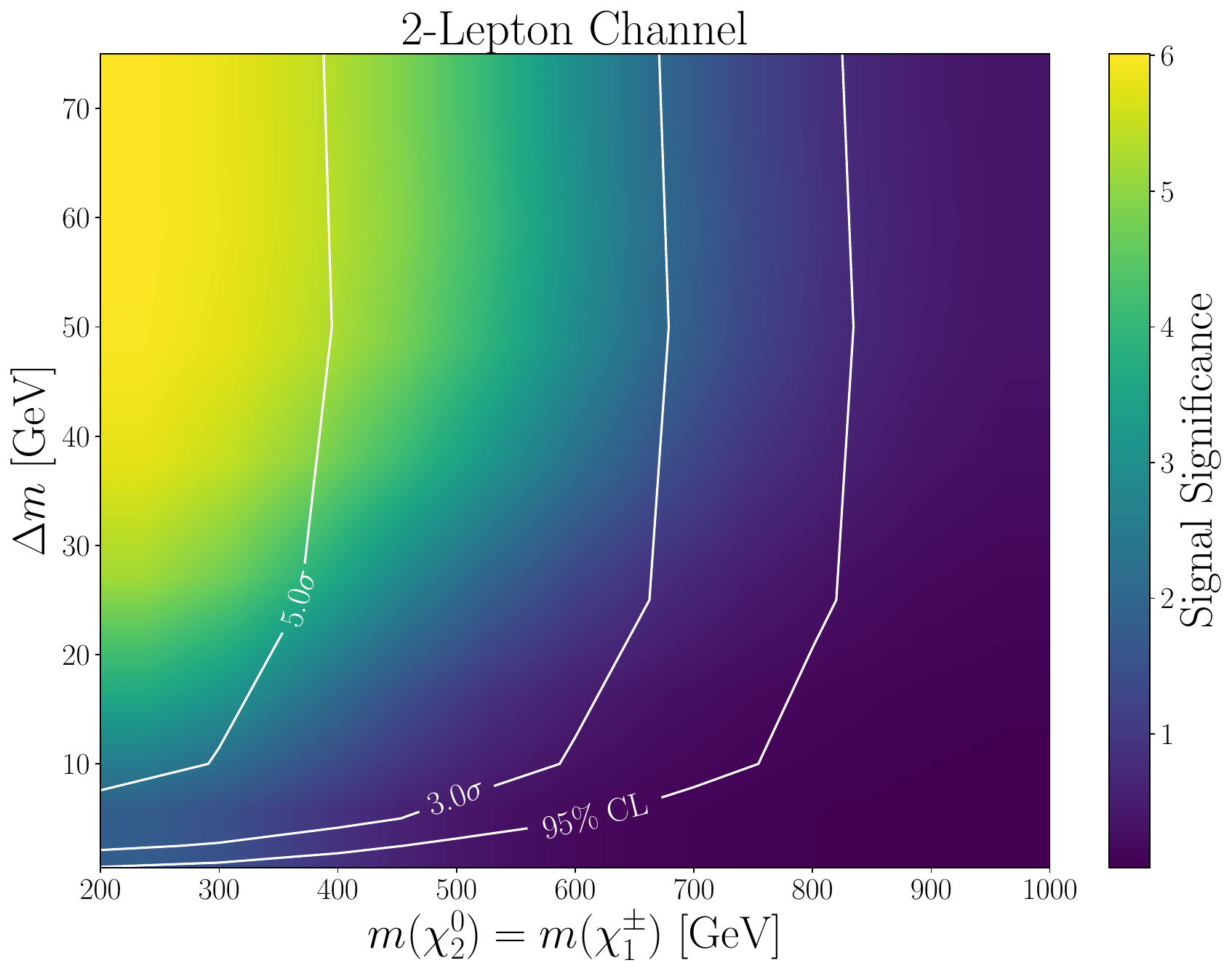}
\end{minipage}%

\begin{minipage}{.5\textwidth}
  \centering
  \includegraphics[width=\linewidth]{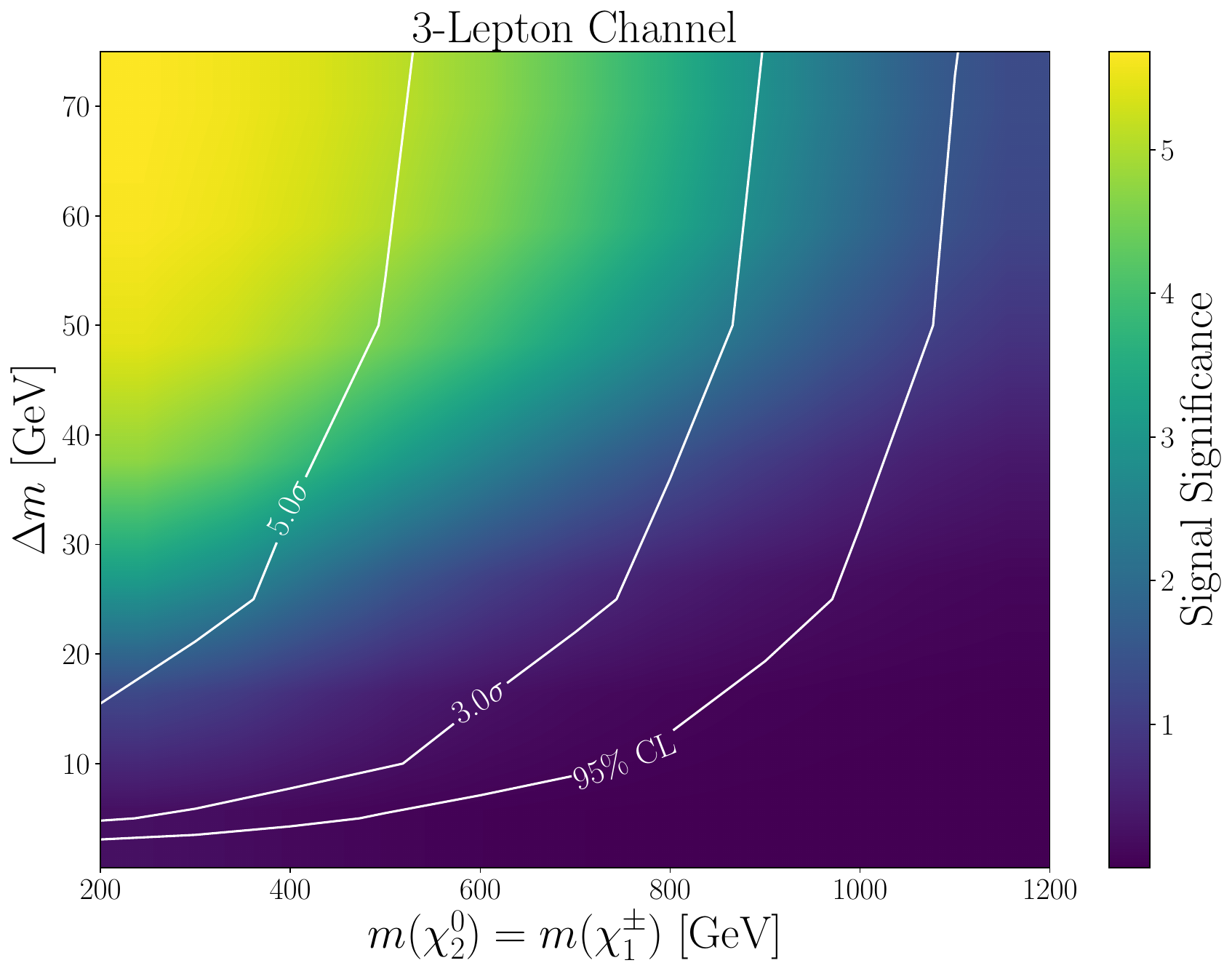}
\end{minipage}%
\begin{minipage}{.5\textwidth}
  \centering
  \includegraphics[width=\linewidth]{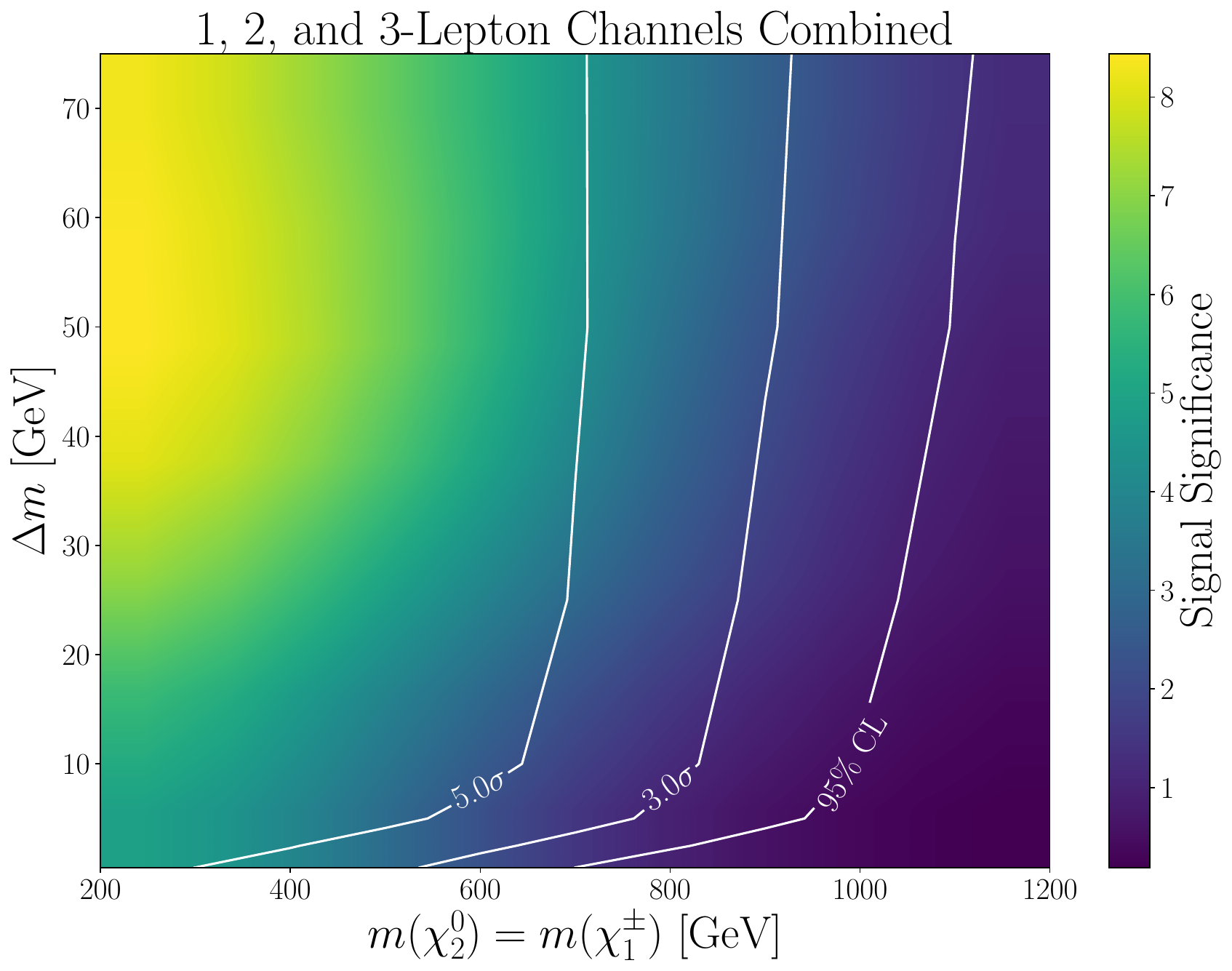}
\end{minipage}%
    \caption{Projected signal significance contours for the 1-lepton (top-left), 2-lepton (top-right), 3-lepton (bottom-left), and combined 1, 2, and 3-lepton (bottom-right) channels as a function of the $\ch$ mass and the mass difference $\Delta m$ for the Wino-Bino light-slepton scenario assuming an integrated luminosity of 3000 $\mathrm{fb}^{-1}$ corresponding to the HL-LHC.}
    \label{fig:SSLS3000}
\end{figure*}

% \begin{figure*}
%     \centering
%     \begin{minipage}{.5\textwidth}
%   \centering
%   \includegraphics[width=\linewidth]{figures/SS_3000_1lep_higgs.pdf}
% \end{minipage}%
% \begin{minipage}{.5\textwidth}
%   \centering
%   \includegraphics[width=\linewidth]{figures/SS_3000_2lep_higgs.pdf}
% \end{minipage}%

% \begin{minipage}{.5\textwidth}
%   \centering
%   \includegraphics[width=\linewidth]{figures/SS_3000_3lep_higgs.pdf}
% \end{minipage}%
% \begin{minipage}{.5\textwidth}
%   \centering
%   \includegraphics[width=\linewidth]{figures/SS_3000_combined_higgs.pdf}
% \end{minipage}%
%     \caption{Projected signal significance contours for the 1-lepton (top-left), 2-lepton (top-right), 3-lepton (bottom-left), and combined 1, 2, and 3-lepton (bottom-right) channels as a function of the $\ch$ mass and the mass difference $\Delta m$ for the higgsino-like LSP scenario assuming an integrated luminosity of 3000 $\mathrm{fb}^{-1}$ corresponding to the HL-LHC.}
%     \label{fig:SSHGS3000}
% \end{figure*}

\begin{figure*}
    \centering
    \begin{minipage}{.5\textwidth}
  \centering
  \includegraphics[width=\linewidth]{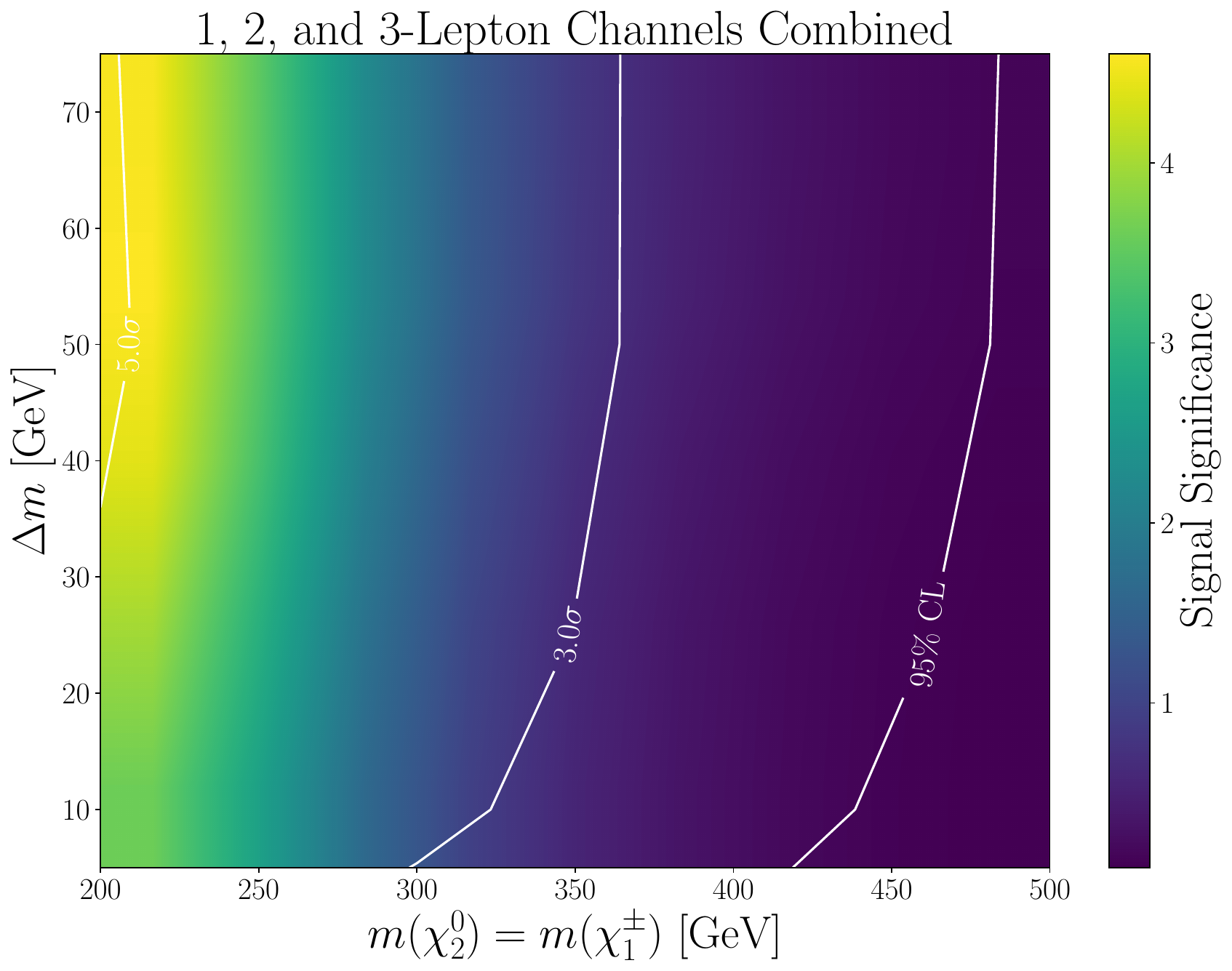}
\end{minipage}%
\begin{minipage}{.5\textwidth}
  \centering
  \includegraphics[width=\linewidth]{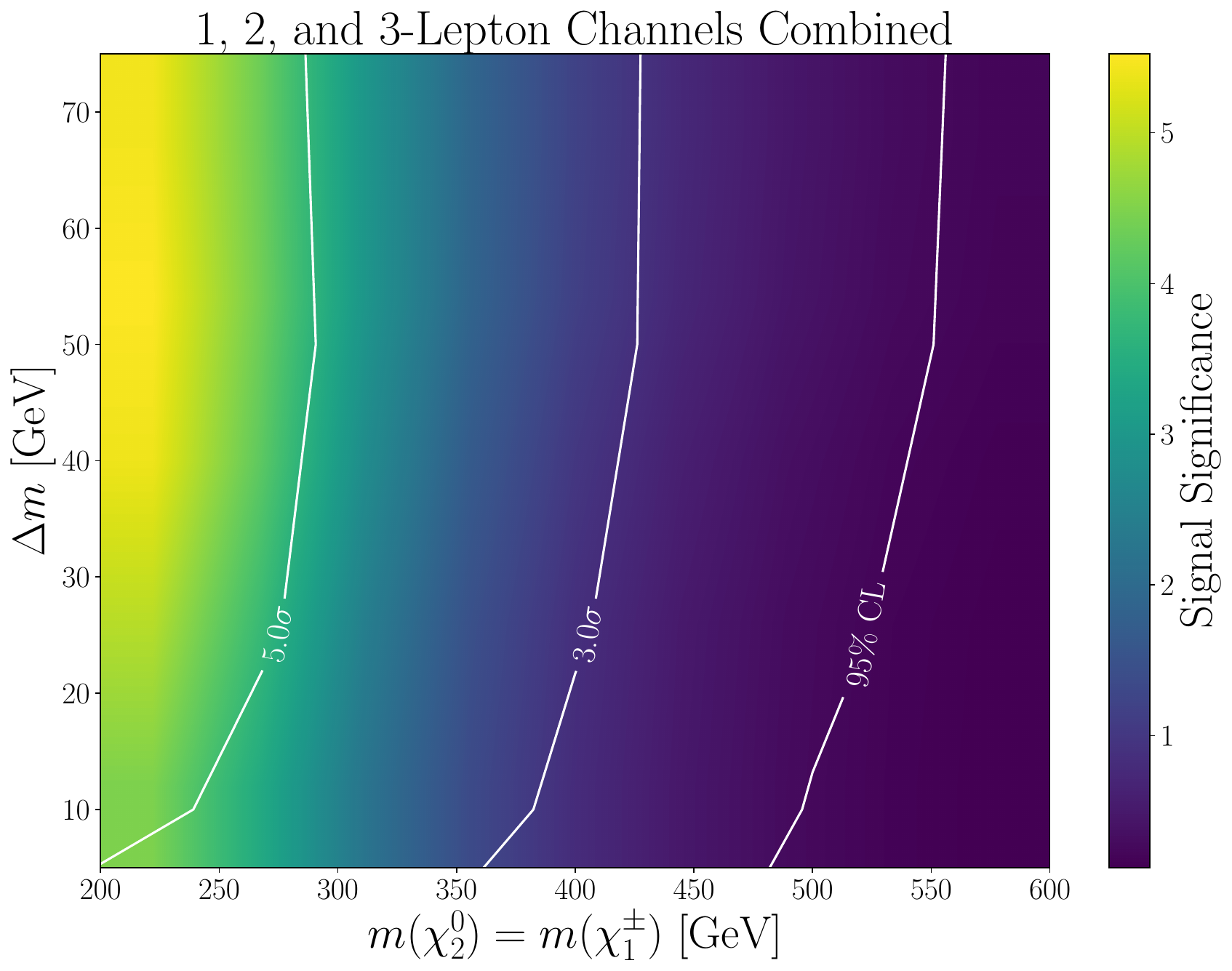}
\end{minipage}%
    \caption{Projected signal significance contours corresponding to integrated luminosities of 137 $\mathrm{fb}^{-1}$ (left) and 300 $\mathrm{fb}^{-1}$ (right) for the combined 1, 2, and 3-lepton channels as a function of the $\ch$ mass and the mass difference $\Delta m$ for the Wino-Bino $W^*/Z^*$ scenario.}
    \label{fig:SSBW137300}
\end{figure*}

\begin{figure*}
    \centering
    \begin{minipage}{.5\textwidth}
  \centering
  \includegraphics[width=\linewidth]{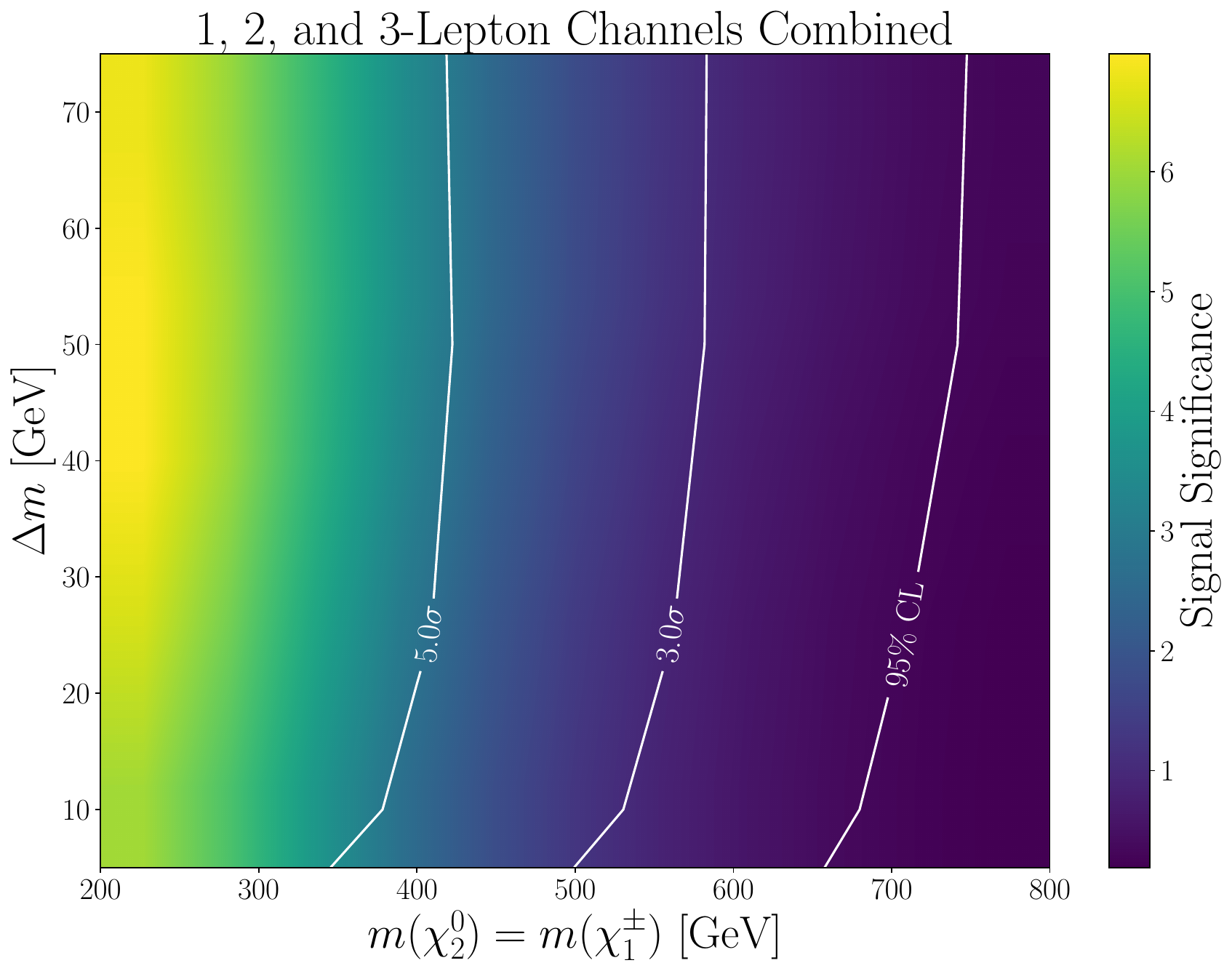}
\end{minipage}%
\begin{minipage}{.5\textwidth}
  \centering
  \includegraphics[width=\linewidth]{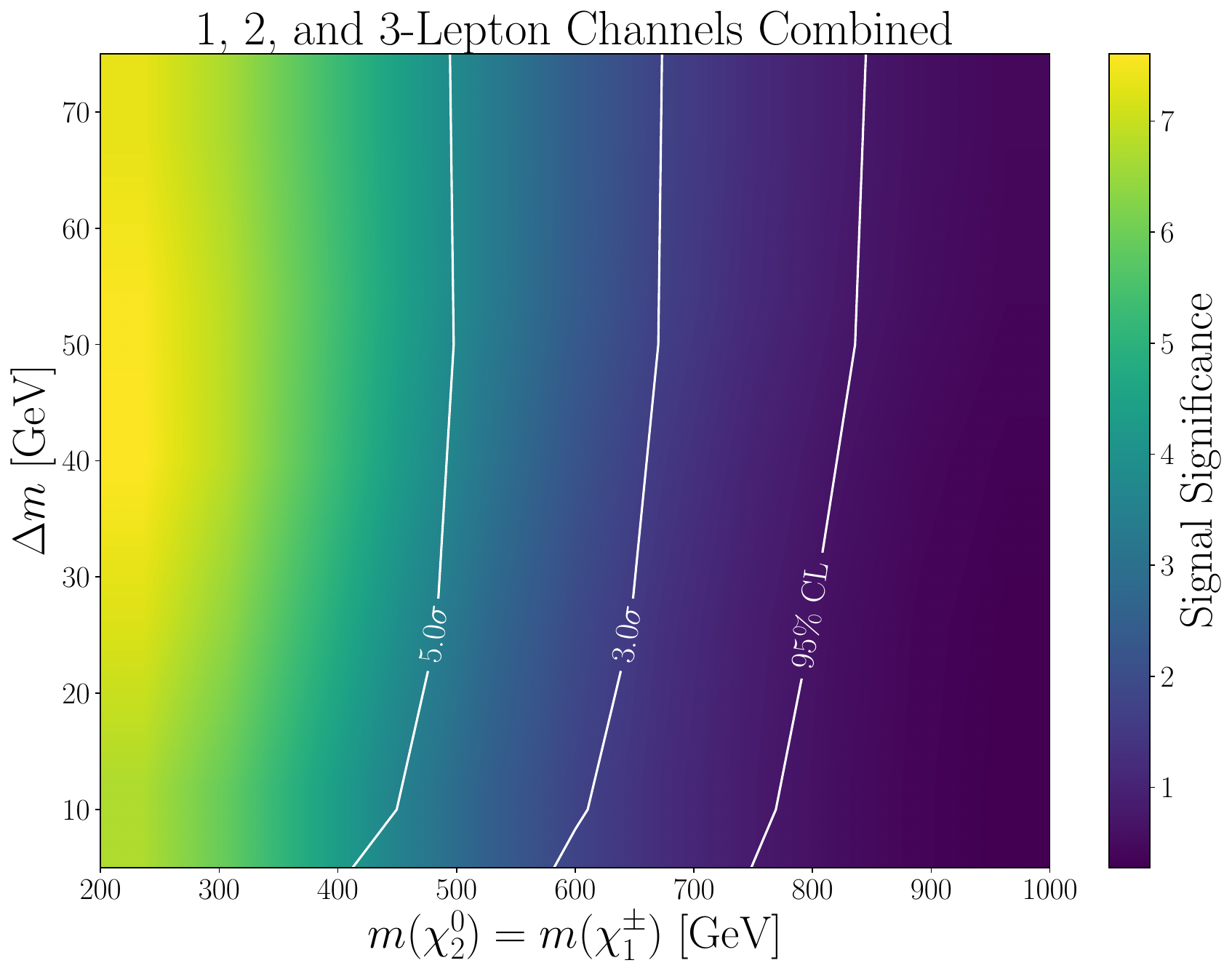}
\end{minipage}%
    \caption{Projected signal significance contours corresponding to integrated luminosities of 137 $\mathrm{fb}^{-1}$ (left) and 300 $\mathrm{fb}^{-1}$ (right) for the combined 1, 2, and 3-lepton channels as a function of the $\ch$ mass and the mass difference $\Delta m$ for the Wino-Bino light-slepton scenario.}
    \label{fig:SSLS137300}
\end{figure*}

% \begin{figure*}
%     \centering
%     \begin{minipage}{.5\textwidth}
%   \centering
%   \includegraphics[width=\linewidth]{figures/SS_137_combined_higgs.pdf}
% \end{minipage}%
% \begin{minipage}{.5\textwidth}
%   \centering
%   \includegraphics[width=\linewidth]{figures/SS_300_combined_higgs.pdf}
% \end{minipage}%
%     \caption{Projected signal significance contours corresponding to integrated luminosities of 137 $\mathrm{fb}^{-1}$ (left) and 300 $\mathrm{fb}^{-1}$ (right) for the combined 1, 2, and 3-lepton channels as a function of the $\ch$ mass and the mass difference $\Delta m$ for the higgsino-like LSP scenario.}
%     \label{fig:SSHGS137300}
% \end{figure*}

Figures \ref{fig:SSBW3000}, \ref{fig:SSLS3000}, and \ref{fig:SSHGS3000}, show the expected signal significance (on the $z$-axis), as a function of the $\neut$ and $\ch$ masses on the $xy$-plane for the Wino-Bino $Z^*/W^*$ and light slepton scenarios respectively. For these plots, we assume an integrated luminosity of $3000~\mathrm{fb}^{-1}$ expected by the end of the high-luminosity LHC era. The white contours delimit the $5\sigma$ discovery region, $3\sigma$ excess contour, and the projected $95\%$ confidence level (CL) exclusion contour if there is no statistically significant evidence of an excess. 

In the top-left panels of each figure, we display the results from the single-lepton channel. For the Wino-Bino $W^{*}/Z^{*}$ scenario, there is $5\sigma$ discovery potential for $m(\ch)$ up to about 310 GeV and $\Delta m = 10$ GeV, and $m(\ch) < 250$ (200) GeV for $\Delta m = 2.5$ (30) GeV. The $3\sigma$ sigma excess contour extends to about $m(\ch)\approx 600$ GeV for $\Delta m \approx 5$--10 GeV, but the single-lepton channel loses sensitivity for higher $\Delta m$, reducing its $3\sigma$ sensitivity to about 350 GeV for $\Delta m \approx 75$ GeV. 
%; $m(\ch)$ up to $250$ GeV with $\Delta m < 10$ GeV for the light slepton scenario; $m(\ch)$ up to $350$ GeV with $\Delta m < 15$ GeV for the higgsino-like LSP scenario. 
The same trend is observed for the Wino-Bino light-slepton scenario, where the $5\sigma$ contours extend to $m(\ch) < 400$ GeV for $\Delta m = 10$ GeV, and $m(\ch) < 325$ (300) GeV for $\Delta m = 2.5$ (30) GeV. For the light-slepton scenario, the $3\sigma$ contours extend to $m(\ch)\approx 425$ GeV at $\Delta m \approx75$ GeV. Meanwhile, the expected $95\%$ CL exclusion bounds for the Wino-Bino $Z^*/W^*$ and light-slepton scenarios are $m(\ch)\approx 750 (475)$ GeV for $\Delta m < 5 (75)$ GeV and $m(\ch)\approx 900 (575)$ GeV for $\Delta m < 5 (75)$ GeV, respectively. The panels demonstrate that the VBF single-lepton channel is an effective probe for small $\Delta m$ values near 10 GeV, and rapidly lose sensitivity for increasing $\Delta m$ values. This phenomenon is observed since the average momenta of the leptons become larger with increasing $\Delta m$ values, therefore, events with two or three leptons are selected more preferably.

The top-right panels in Figures \ref{fig:SSBW3000} and~\ref{fig:SSLS3000} show the expected signal significance for the dilepton channel. The projected 95\% CL exclusion region reaches $m(\ch)\approx700$ (800) GeV for $\Delta m >25$ GeV in the Wino-Bino $Z^*/W^*$ (light slepton) scenario. The $3\sigma$ region extends up to $m(\ch)\approx 550$ (650) GeV for $\Delta m >25$ GeV in the $Z^*/W^*$ (light slepton) scenario. In terms of discovery reach, the $5\sigma$ sensitivity extends up to $m(\ch) \approx 325$ GeV for $\Delta m =20$ GeV in the $Z^*/W^*$ scenario, and $m(\ch)\approx 375$ GeV for $\Delta m >25$ GeV for the light-slepton scenario. However, as expected, the 2-lepton channel does not provide discovery potential in any model for $\Delta m < 7.5$ GeV. This phenomenon occurs because the average momentum of the leptons decreases as the $\Delta m$ values decrease, making events with a single lepton more likely to be selected.

The 3-lepton channel, displayed in the bottom-left panels of Figures \ref{fig:SSBW3000} and~\ref{fig:SSLS3000}, improves the exclusion and discovery reach for higher values of $\Delta m$. This is expected since the average momentum of the leptons increases as the $\Delta m$ values increase, making the 3-lepton signal acceptance larger than 1-lepton and 2-lepton signal acceptances, as is highlighted by Table~\ref{tab:sel-eff}. However, the signal significance drops considerably with decreasing $\Delta m$. The projected 95\% CL exclusion region extends to $m(\ch) \approx \{ 350,700,925 \}$ $(\{ 400,850,1100 \})$ for $\Delta m = \{ 5, 15, 75 \}$ GeV for the $Z^*/W^*$ (light-slepton) scenario. Meanwhile, the 3$\sigma$ region extends to $m(\ch) \approx 725$ (850) GeV for $\Delta m = 75$ GeV for the $Z^*/W^*$ (light-slepton) scenario.

Combining the three search channels allows us to have sensitivity to a broader range of $\{ m(\ch),\Delta m \}$ phase space. The signal significance for the combination of the 1-, 2-, and 3-lepton channels is shown in the bottom-right panel of Figures \ref{fig:SSBW3000} and~\ref{fig:SSLS3000}. For the Wino-Bino $W^*/Z^*$ scenario, the projected 95\% CL exclusion region extends up to $m(\ch) \approx 950$ GeV for $\Delta m > 50$ GeV, and then dips gradually to $m(\ch) \approx 750$ GeV for $\Delta m \approx 5$ GeV. The 3$\sigma$ region covers $m(\ch) < 750$ GeV for $\Delta m > 50$ GeV, and then again dips gradually to $m(\ch) \approx 625$ GeV for $\Delta m = 5$ GeV. The $5\sigma$ discovery region goes up to $m(\ch)\approx 550$ GeV for $\Delta m > 25$ GeV then dips to $m(\ch)\approx 420 $ GeV for $\Delta m\approx 5$ GeV. In the Wino-Bino light-slepton scenario, the projected 95\% CL exclusion region extends up to $m(\ch) \approx 1100$ GeV for $\Delta m > 50$ GeV, gradually decreasing to $m(\ch) \approx 900$ GeV when $\Delta m \approx 5$ GeV. The $3\sigma$ region reaches $m(\ch) \approx 875$ GeV for $\Delta m > 40$ GeV, and then gradually drops to $m(\ch) \approx 700$ GeV as $\Delta m$ approaches 5 GeV. The $5\sigma$ discovery region extends to $m(\ch) \approx 650$ GeV for $\Delta m > 25$ GeV, dipping to $m(\ch) \approx 500$ GeV for $\Delta m \approx 5$ GeV.

%Finally, for the higgsino-like LSP scenario, the predicted 95\% CL exclusion boundary reaches beyond $m(\ch) \approx 1200$ GeV for $\Delta m > 60$ GeV, tapering off to $m(\ch) \approx 1100$ GeV as $\Delta m$ reduces to 5 GeV. The $3\sigma$ region extends to $m(\ch) \approx 990$ GeV for $\Delta m > 35$ GeV, gradually reducing to $m(\ch) \approx 850$ GeV at $\Delta m \approx 5$ GeV. The $5\sigma$ discovery region is projected to go up to $m(\ch) \approx 600$ GeV for $\Delta m > 25$ GeV, decreasing to $m(\ch) \approx 400$ GeV for $\Delta m \approx 5$ GeV.

In Figures \ref{fig:SSBW137300} and \ref{fig:SSLS137300} we show the combined signal significance for integrated luminites of $137~\mathrm{fb}^{-1}$ (left panels) and $300~\mathrm{fb}^{-1}$ (right panels). For the Wino-Bino $W^*/Z^*$ scenario, the projected 95\% CL exclusion region extends up to $m(\ch) \approx 480~(550)$ GeV for $\Delta m > 50$ GeV and then dips gradually to reach $m(\ch) \approx 420~(470)$ GeV for $\Delta m \approx 5$ GeV for $\mathcal{L}_{\mathrm{int}}=137~(300)~\mathrm{fb}^{-1}$. In the Wino-Bino light-slepton case, the projected 95\% CL exclusion region goes to $m(\ch) \approx 750~(825)$ GeV for $\Delta m > 50$ GeV and then dips gradually to reach $m(\ch) \approx 650~(750)$ GeV for $\Delta m \approx 5$ GeV for $\mathcal{L}_{\mathrm{int}}=137~(300)~\mathrm{fb}^{-1}$. Therefore, the expected discovery and exclusion reach using our unique VBF multi-lepton topology, combined with a novel TabNet deep-learning model, can be extended to a regime of the compressed mass spectrum parameter space that is unconstrained by current searches at the LHC.

\section{Discussion}\label{sec:discussion}

In this work, we assess the discovery potential of using VBF topologies in conjunction with machine learning techniques to probe electroweakino production in the compressed-mass spectrum regions. The results presented indicate that the use of soft lepton final states containing VBF-tagged jets, in combination with a novel sequential attention-based machine learning classifier, improves signal-background discrimination, significantly enhancing the discovery potential and sensitivity reach of current and future LHC compressed SUSY electroweakino searches into previously inaccessible regions of the parameter space.

We also study the mixing effects between pure electroweak and mixed QED-QCD diagrams which can contribute to the production of electroweakino pair production and two jets (i.e., $\widetilde{\chi}_{m}\widetilde{\chi}_{n}jj$). . The interference between these diagrams can affect the total cross-section and the kinematics of interest for these studies, and thus to study this effect, we introduce the RKI parameter, which simultaneously quantifies the effect on the signal production cross section and the kinematic variable of interest. We find that the interference effect depends on both $m(\ch)$ and $\Delta m$ and can be up to about 20\% for the $\{ m(\ch), \Delta m \}$ values and ``ino'' composition considered
in this paper. Understanding these interference effects is essential for accurate training of the TabNet deep-learning model in order to maximize discovery reach, and should there exist evidence of new physics, is also important to correctly extract measurements of the SUSY parameters. 

Traditional SUSY searches have been limited by the difficulty in reconstructing low-momentum particles in compressed-mass spectra, where the mass differences between the lightest supersymmetric particle (LSP) and other supersymmetric particles are small, and using them effectively to suppress SM backgrounds. Our study demonstrates that the VBF topology provides a powerful means to boost these soft final-state particles, which, in conjunction with our novel machine learning approach, facilitates their identification with enhanced background rejection power. We also point out that the use of a novel attention-based deep learning architecture enables us to achieve better signal significance than traditional models such as Multi-Layer Perceptrons, while also having the interpretability of Boosted Decision Trees.

The combination of a final state with soft leptons, VBF-tagged jets, and machine learning techniques can allow us to set exclusion limits at the 95\% confidence level for electroweakinos in the Wino-Bino $W^{*}/Z^{*}$ and light-slepton scenarios well beyond the capabilities of current searches at ATLAS and CMS, with projected exclusion boundaries extending up to $m(\ch)\approx 1.1 $ TeV for $50 < \Delta m < 75$ GeV, and up to $m(\ch)\approx 600$ GeV for $\Delta m \approx 1$--2 GeV. We note that although experimental triggers for the HL-LHC are beyond the scope of this work, the proposed analysis strategy in this paper requires a suitable soft lepton plus VBF trigger so that the full spectrum of the missing transverse momentum can be used as an input variable to the machine learning algorithm. In conclusion, we strongly encourage the ATLAS and CMS Collaborations to consider the proposed analysis strategy.
%conclusion, this work demonstrates that the combination of VBF processes with advanced machine learning techniques provides a promising pathway for probing electroweak SUSY particles in challenging compressed-mass spectrum regions. As the LHC and HL-LHC continue to collect data, the general methodology outlined in this paper can be a key tool in the search for new physics.

\begin{acknowledgements}
A. F. would like to thank the constant and enduring financial support received for this project from the faculty of science at Universidad de Los Andes (Bogot\'a, Colombia) through the projects INV-2023-178-2999 and INV-2023-175-2957. A. G and U. Q. acknowledge the funding received from the Physics \& Astronomy department at Vanderbilt University and the US National Science Foundation. 
This work is supported in part by NSF Award PHY-2111554, a Vanderbilt Undergraduate Research Fellowship, and a Vanderbilt Seeding Success Grant.
\end{acknowledgements}

\bibliography{apssamp}% Produces the bibliography via BibTeX.

\end{document}